\documentclass[a4paper]{jpconf}
\pdfoutput=1
\usepackage{amsmath}
\usepackage[utf8]{inputenc}
\usepackage[english]{babel}
\usepackage{multirow}
\usepackage{graphicx}
\usepackage[numbers,sort&compress]{natbib}
\usepackage[usenames,dvipsnames,x11names]{xcolor}
\usepackage{hyperref}
\usepackage{txfonts}
\usepackage[T1]{fontenc}

\begin{document}
	
\title{A Monte Carlo method for solving the electromagnetic scattering problem in dielectric bodies}

\author{Hector Lopez-Menchon (hector.lopez@upc.ed)\\ Juan M. Rius (juan-manuel.rius@upc.edu)\\Alexander Heldring (heldring@tsc.upc.edu)\\Eduard Ubeda (eduard.ubeda@upc.edu)}

\address{Signal Theory and Communications Department, UPC}

\begin{abstract}
	In this work, we develop a novel Monte Carlo method for solving the electromagnetic scattering problem. The method is based on a formal solution of the scattering problem as a modified Born series whose coefficients are found by a conformal transformation. The terms of the Born series are approximated by sampling random elements of its matrix representation, computed by the Method of Moments. Unlike other techniques as the Fast Multiple Method, this Monte Carlo method does not require communications between processors, which makes it suitable for large parallel executions. 
\end{abstract}

\section{Introduction}

\subsection{Born series}

The Born series has been widely used to solve scattering problems. It was first introduced by Max Born in the 1920s for solving the Lippmann-Schwinger equation in the framework of the quantum theory of scattering \cite{zettili_quantum_2009}. It is also used to solve electromagnetic \cite{mishchenko_libro_gordo}, acoustic \cite{Kouri} or seismic scattering problems. The Born series mathematically equivalent to the Neumann or Liouville-Neumann series, which was proposed in the 19th century as a method for solving a certain kind of functional equations. 

The Neumann series allows to expand the inverse of a linear operation as an operator power series. Since the coefficients of the power series are known beforehand, the Neumann series is closely related fo fixed-point iteration methods \cite{hutson_applications_1980}. For an equation of the form $x=f(x)$, fixed-point iteration methods start from an initial guess $x_0$ and iteratively refine it as $x_k = f(x_{k-1})$. The Neumann series is equivalent to a fixed-point iteration method when $f(x)$ is a linear operator. 

When applied to solve scattering problems, the Neumann series receives the name of Born series. A large variety of methods for electromangetic scattering rely on a Born series formalism. Its most classical form consist on a Neumann series directly built on the Volume Integral Equation \cite{gbur_2011}. A common variation is the Generalized Born Series (GBS) \cite{schuster_hybrid_1985,martin_multiple_2006}, which is usually applied to sets of electromagnetic scatterers. In the GBS, the local scatering problems associated to each object are solved by an arbitrary method (MoM, FDTD, Mie, or others), and the global problem involving interactions between scattering objects is addressed with a Born series. So, the GBS can be thought as a domain decomposition method \cite{bourlier_sub-domain_2015}. The Neumann series is also used in the framework of radiative transfer \cite[p.~65]{mishchenko_libro_gordo}. In the radiative transfer framework, ensembles of a very large number of particles are considered, and each particle is assumed to be in the far field region of all the other particles in the system \cite{mishchenko_independent_2018}. The local scattering problem is solved by a far-field analytic method, and the overall field is computed by a fixed-point iterative method known as order of scattering expansion \cite{mishchenko_independent_2018}, multiple scattering \cite{mishchenko_multiple_2008} or successive order of interaction \cite{heidinger_successive-order--interaction_2006}. This can be understood as a Neumann series applied to the far-field Foldy-Lax equations \cite{mishchenko_libro_gordo} or as a particular case of GBS. Another method that relies on a Neumann or Born series formulation is Iterative Physical Optics (IPO) \cite{burkholder_forward-backward_2005,gershenzon_paper_2018,obelleiro_1995}. The IPO method recursively applies the Physical Optics approximation to account for high-frequency multiple interactions in problems like the computation of Radar Cross Section from large targets or antennas radiating in complex environments. 

All the methods pointed above employ a Neumann series formalism to invert a linear operator. However, the Neumann series suffers from serious convergence limitations, which restricts the range of situations where it can be applied. An equation of the form $u = Hu +c$, where $H$ is a linear operator, only can be solved by a Neumann series if the the spectral radius of $H$ is smaller than one \cite{renardy_introduction_2004}. For this reason, we developed a modified Born series with enhanced convergence \cite{menchon-born}. In this work, we propose a particular implementation of this method where the coefficients are obtained by solving a conformal mapping problem. 

\subsection{Monte Carlo methods in EM scattering} 

Monte Carlo (MC) methods are widely used in computational physics to find approximate solutions of problems where exact methods are impossible to apply of prohibitively costly. They are also applied in computational electromagnetics for solving electrostatic or electrodynamic problems \cite{sadiku_monte_nodate}. In the case of multiple scattering, MC methods are a fundamental tool in radiative transfer \cite{noebauer_monte_2019,barker_monte_2003,deutschmann_monte_2011}, but are seldom applied to more general scattering problems. The reason is that Monte Carlo methods require the solution of the problem to be written as a formal expression that can be approximated by random sampling. This is very difficult in many scattering problems. In the particular case of radiative transfer, the solution of the scattering problem can be written as  Neumann series whose result is approximated by a random sampling method which is equivalent to the classical Ulam-Neumann Monte Carlo method for solving linear system of equations. The Ulam-Neumann method does not only require the Neumann series to be convergent, but the operator should also fulfill additional conditions \cite{ji_convergence_2013}. For radiative transfer problems, the interactions between scattering objects are weak. Thus, the scattering operator fulfills the conditions for both the Neumann series \cite{deutschmann_monte_2011} and the Ulam-Neumann method to converge.

As we will see in Section \ref{sec:numerical_experiments}, the Ulam-Neumann method cannot be applied to the cases of interest here. However, independently sampling the elements of the Born series allows to draw a Monte Carlo solution of the scattering problem.

\subsection{Objectives of this work}

In this work, we develop a Monte Carlo method based on a modified Born series. Unlike the Ulam-Neumann, our method approximates the different terms of the Born series in an independent way. Our method can be understood as a set of multidimensional integrals approximated by Monte Carlo integration. The restrictive convergence conditions of the Ulam-Neumann algorithm no longer apply, since the only convergence condition is that the integrand should be bounded. In this way, our Monte Carlo method can be applied to a large set of scattering problems. 

This work is organized as follows. In Section \ref{sec:theory} we give the theoretical foundations of our method. Section \ref{sec:conformal_maps} describes the necessary basics of conformal mapping and how it is applied to solve linear systems of equations. Section \ref{sec:EM_problem} describes the electromagnetic scattering problem in the Volume Integral Equation formulation, and in Section \ref{sec:EM_MC} we present the Monte Carlo method for solving it. Section \ref{sec:numerical_experiments} shows some numerical results. Section \ref{sec:scalability} is devoted to a scalability model that predicts how the Monte Carlo method would scale in a large parallel machine depending on the network topology, and compares it with other classical method. In Section \ref{sec:conclusions} we expose the conclusions. 

\section{Theory}\label{sec:theory}

\subsection{Conformal Maps}\label{sec:conformal_maps}

\subsubsection{General Concepts}

Let $U$ and $V$ be open subsets in the complex plane $\mathbb{C}$. A bijective holomorphic function $f \colon U \to V$ is called a conformal map \cite[p.~206]{stein_complex_2003}. From this, several interesting properties of $f$ follow: The inverse function $g \colon V \to U$ exists and is also holomorphic \cite[p.~208]{stein_complex_2003}. Also, since $f$ is injective and holomorphic in $U$, then $f'(z) \neq 0$ for all $z \in U$ \cite[p.~206]{stein_complex_2003}\cite[p.~209]{lang_complex_1999} (the converse is not necessarily true). For the same reason, $g'(z)\neq0$ for all $z \in V$. Another useful result is that a composition of conformal maps is also a conformal map \cite[p.~209]{lang_complex_1999}.

The above definition of a conformal map is not universally accepted. For some authors, a map $f \colon U \to V$ is conformal if it is holomorphic and $f'(z)\neq0$ for all $z \in U$. A map satisfying these conditions locally preserves angles, which matches the historical meaning of the world \textit{conformal}. This second definition is less restrictive, since it does not require $f$ to be a bijection between $U$ and $V$. For example, the function $f(z)=z^2$ on  (este on esta bien, mirado de Stein) $\mathbb{C}-\{0\}$ is holomorphic and fulfills $f'(z)\neq0$, but is not bijective on its domain (although it is locally bijective). Then, this function would be a conformal map according to the latter definition, but not according to the former one. We will use the former definition. 

If $U$ and $V$ are simply connected, the Riemann Mapping Theorem grants the existence of a conformal map between them \cite[p.~32]{kythe_handbook_2019}. Let $D$ be the unit disk $D=\{z\in\mathbb{C}\colon|z|<1\}$. The Riemann Mapping Theorem states that, \textit{if $U$ is a simply connected open set which is not the whole plane, then there exists a conformal map from $U$ to the unit disk}. In order to prove the existence o a conformal map from $U$ to $V$, let $h\colon U \to D$ be a conformal map from $U$ to $D$, and let $s\colon V\to D$ be a conformal map from $V$ to $D$. The existence of these functions is granted by the Riemann Mapping Theorem. The inverse function $s^{-1}\colon D \to V$ exists and is also a conformal map. Since the composition of two conformal maps is also a conformal map, then there is a conformal map $s^{-1}\circ h$ from $U$ to $V$. Fig. [alguna] shows a schematic representation of this.

The Riemann Mapping Theorem also sates that the mapping $h$ from $U$ to $D$ is essentially unique: it is uniquely determined provided that $f(z_0)=0$ for $z_0\in U$. 

The Riemann Mapping Theorem also allows to draw results about the behavior of the boundaries of the open subsets. In particular, we would like to know if the map $f \colon U \to V$ extends to a bijection from the boundary of the domain region to the boundary of the image region. Let $\partial U$ be the closure of $U$, and $\partial V$ the closure of $V$. If $\partial U$ and $\partial V$ are regular, then $f$ extends to a bijection $f \colon \partial U \to \partial V$

The following proposition holds \cite[p.~30]{lang_complex_1999}: Let $h\colon U \to D$ a conformal map (analytic isomorphism in original). Let $\gamma$ be a proper analytic arc contained in the boundary of $U$ and such that $U$ lies on one side of $\gamma$. Then $g$ extends to an analytic isomorphism on $U \cup \gamma$.

Thanks to the Schwarz reflection principle \cite{lang_complex_1999} , it is also possible to obtain some results about the behavior of the boundaries of the open domains considered here when a conformal map is applied. Let us consider again the conformal map $h$ from an open, simply connected subset $U$ with boundary $\partial U$ to the open disk $D$ with boundary $\partial D$. The closures \cite[p.~6]{stein_complex_2003} of these sets are $\overline{U}=U \cup \partial U$ and $\overline{D} = D \cup \partial D$ for $U$ and $D$, respectively. Then, if $\partial U$ is piecewise smooth, the conformal map $h$ extends to an analytic bijection \cite[p.~301]{lang_complex_1999}from $\overline{U}$ to $\overline{D}$\cite{stein_complex_2003}. By using the Riemann Mapping Theorem, it is trivial to prove that $f\colon U \to V$ extends to a bijection between the closures $\overline{U}$ and $\overline{V}$ provided that $\partial{U}$ and $\partial{V}$ are piecewise regular. This result about continuous extension can be formulated for more general boundaries \cite{lang_complex_1999}, but this will not be necessary for the cases of interest here. 

\subsubsection{Internal and External Maps}
\label{sec:internal_external}

We will be specially interested in exterior maps. Instead of the complex plane $\mathbb{C}$, we will consider the extended complex plane $\overline{\mathbb{C}}$ (or Riemann sphere) \cite[p.~101]{asmar_complex_2018}, formed the the complex plane plus a point at infinity. Let $\Omega$ be an open, simply connected open set with piecewise regular boundary $\partial \Omega$. The complement of $\Omega$ with respect to the extended complex plane, noted as $\overline{\mathbb{C}} \setminus \Omega$, is simply connected \cite{starke_hybrid_1993,driscoll_schwarz-christoffel_2002}, and also the complement of the unit disk $\overline{\mathbb{C}} \setminus D$. Then, according to the Riemann  Mapping Theorem, there exists a conformal map $\Phi \colon \overline{\mathbb{C}} \setminus \Omega \to \overline{\mathbb{C}} \setminus D$ from the exterior of the set $\Omega$ to the exterior of the unit disk.

\subsubsection{Series expansion of conformal maps}
\label{sec:expansion}

Conformal maps are analytic functions, hence they admit power series expansions. For reasons that will be clear later, we are interested in the power series expansion of $\Phi^{-1} \colon \overline{\mathbb{C}} \setminus D \to \overline{\mathbb{C}} \setminus \Omega$. For this, we start considering the conformal map $T \colon D \to \overline{\mathbb{C}} \setminus \Omega$, depicted in Fig. (). It is clear the conformal map from $D$ to $\overline{\mathbb{C}} \setminus D$ is simply $1/z$. Then, since the exterior map fulfills $\Phi(\infty)=\infty$, $T$ fulfills $T(0)=\infty$. In other words, $T$ presents a singularity at zero. As a consequence of this singularity, a Laurent series (a power series involving negative terms) is necessary to represent $T$ in $D\setminus \{0\}$. Besides, the singularity is of order $1$ \cite{laura-conformal-mapping}, so the only non-zero negative term is the one of order $-1$. Finally, we have the expansion
\begin{equation}\label{eq:first_expansion}
T(z) = C \frac{1}{z}+b_0 + b_1 z^2+ b_2 z^2+ \dots
\end{equation}
which is convergent on the region $D\setminus \{0\}$.

As we said, our objective is to obtain a power series expansion for $\Phi^{-1}$. The map $\Phi^{-1}$ can be represented as a composition of the reciprocal function $1/z$ and $T$. Then,
\begin{equation}\label{eq:laurent_expansion}
\Phi^{-1}(z) = T\left(\frac{1}{z}\right) = C z + b_0 + b_1 \frac{1}{z} + b_2 \frac{1}{z^2} + \dots 
\end{equation}

Some numerical methods for approximating conformal maps use Laurent expansions \cite[p.~95]{laura-conformal-mapping}. Usually, the coefficients are determined by imposing that a known bijection between the boundaries $\partial \Omega$ and $\partial D$ is satisfied. The most straightforward way to find the coefficients is directly employing the Cauchy integral formula. Another classical technique is the method of simultaneous equations of Kantorovich \cite[p.~100]{laura-conformal-mapping}, that builds, from the transformation at the boundaries, a system of equations with the coefficients as unknowns. In our case, as we will see in Section \ref{sec:iterative_method}, the coefficients will be derived from an explicit integral representation of the transformation.  

\subsubsection{The Schwarz-Christoffel transformation}

The Schwarz-Christoffel transformation is a conformal map between a simple polygon and the upper half plane, or between a simple polygon and the unit disk. We are particularly interested in the latter case. It is not the purpose of this text to give a thorough description of the Schwarz-Christoffel transformation. We recommend the work \cite{driscoll_schwarz-christoffel_2002} for a detailed explanation.

Let us consider again the maps $\Phi$ and $T$ and the set $\Omega$ from Section \ref{sec:internal_external} and Section \ref{sec:expansion}. In this case the set $\Omega$ is, in particular, a polygon with $n$ vertices $w_1, \dots, w_n$ given in counterclockwise order. We call $z_k = T^{-1}(w_k)$ the \textit{prevertices} of $\Omega$. Also, $\alpha_k$ are the interior angles corresponding to the vertices of $\Omega$. Then, there is a Schwarz-Christoffel map $T \colon D \to \overline{\mathbb{C}} \setminus \Omega$ given by
\begin{equation}\label{eq:SCmap}
T(z) = A - C \int^{z} v^{-2} \prod_{k=1}^{n} \left(1-\frac{v}{z_k}\right)^{1-\alpha_k} d v,  
\end{equation}
where $A$ and $C$ are some normalization constants. Note that $C$ in (\ref{eq:SCmap}) is the same constant as in (\ref{eq:first_expansion}).

Determining the prevertices $z_k$ and the constants $A$ and $C$ is not easy in general. Fortunately, there are excellent software packages for computing Schwarz-Christoffel transformations. We use the SC Toolbox for MATLAB [cita], which is the successor of the FORTRAN package SCPACK. For a given set of vertices $w_k$, the SC Toolbox computes the map $T \colon D \to \overline{\mathbb{C}} \setminus \Omega$, its inverse and the associated parameters. The package does not directly compute the coefficients of (\ref{eq:first_expansion}). However, it allows to compute the so called Faber polynomials \cite{driscoll_schwarz-christoffel_2002}, from which it is trivial to obtain the coefficients of (\ref{eq:first_expansion}). The Faber polynomials $F_k(z)$ satisfy the recurrence relation
\begin{equation}
F_k(z) = \frac{1}{C}\left(z F_{k-1}(z)-(b_0 F_{k-1}(z)+\dots+c_{k-1}F_0(z))-(k-1) b_{k-1}\right)
\end{equation}
for $k\geq 2$. For $k=0,1$ we have $F_0 = 1$ and $F_1(z)=(z-c_0)/C$.

The Faber polynomials are closely linked to the iterative method that we will describe in Section (tal). Indeed, it is sometimes referred as the Faber method. For a detalied discussion about the topic, see [SV93]. 

\subsubsection{The iterative method}\label{sec:iterative_method}

Let us consider the matrix equation 
\begin{equation}\label{eq:system}
u = H u + c, \qquad \qquad u,c \in \mathbb{C}^{n}, \ H \in \mathbb{C}^{n \times n}. 
\end{equation}
Although the matrix case is taken for convenience, the techniques described below can be also applied to general vector spaces where $u$ and $c$ are vectors and $H$ is a linear operator. 

A classical way to solve equations of  the form of (\ref{eq:system}) is the Neumann series. Let $\lambda_i$ be the eigenvalues of $H$. The spectral radius of $H$ is defined as $\rho(H) \equiv \max_i |\lambda_i|$. Then, if $\rho(H)<1$,
\begin{equation}\label{eq:neumann_series}
(I-H)^{-1} = \sum_{k=0}^{\infty}H^k . 
\end{equation}
This expansion is known as the Neumann series\footnote{It is possible to find a less restrictive convergence condition for the Neumann series. If $\|H\|<1$, then (\ref{eq:neumann_series}) converges in the operator norm \cite{renardy_introduction_2004}. Note that, since $|\lambda_i|\leq\|H\|$ for a linear and bounded operator $H$ \cite{hanson-yakovlev}, the condition $\|H\|<1$ implies that $\rho(H)<1$, but the converse is not necessarily true.}. 

The solution of (\ref{eq:system}), $u=(I-H)^{-1}c$, can be approximated by a vector $u_K$, resulting from applying an order $K$ truncation of the Neumann series to $c$:
\begin{equation}\label{eq:truncated_neumann}
u_K \equiv \sum_{k=0}^K H^k c \approx  u.
\end{equation}

The Neumann series can be obtained as the Taylor series expansion of $(I-H)^{-1}$. This, besides the mathematical insight it provides, will allow us to develop a formalism similar to (\ref{eq:truncated_neumann}) that converges when the condition $\rho(H)$ is not fulfilled. First, let us define the resolvent of the operator $H$,
\begin{equation}\label{eq:resolvent}
R_\lambda(H) \equiv (H-\lambda I)^{-1},
\end{equation}
where $\lambda$ is a complex parameter. In particular, we will be interested in evaluating the resolvent at $\lambda=1$. For convenience, let us consider the reciprocal variable $\mu = \lambda^{-1}$. The resolvent (\ref{eq:resolvent}) is then written as
\begin{equation}\label{eq:resolvent_definition}
R_\lambda (H) = (H-\lambda I)^{-1} = -\frac{1}{\lambda} (I-\frac{1}{\lambda}H)^{-1}=-\mu(I-\mu H)^{-1} \equiv \tilde{R}_{\mu}(H)
\end{equation}
Now, we expand the $(I-\mu H)^{-1}$ factor as a Taylor series on the variable $\mu$ around $\mu = 0$
\begin{equation}\label{eq:taylor_resolvent}
\begin{split}
\tilde{R}_\mu(H) & = -\mu (I-\mu H)^{-1} = -\mu \sum_{k=0}^{\infty} \frac{1}{k!}\left.{ \frac{\partial^k}{\partial \mu^k} (I-\mu H)^{-1}} \right|_{\mu=0} \mu ^k \\ 
& = -\mu  \sum_{k=0}^{\infty}  (I-\mu H )^{-k+1} H^k \big|_{\mu=0} \mu^k = -\mu \sum_{k=0}^{\infty} H^k \mu^k 
\end{split}
\end{equation}
It is well known that a Taylor series centered at a point $\mu_0$ is convergent in an open disk where the function is holomorphic \cite{lang_complex_1999}. In the case of (\ref{eq:taylor_resolvent}), the resolvent $\tilde{R}_{\mu}(H)$ is holomorphic if the matrix $I-\mu H$ is not singular. From (\ref{eq:resolvent_definition}), the values $\mu_i$ for which $I-\mu H$ is singular correspond to the values $\lambda_i$ for which $H-\lambda I$ is singular, linked via $\mu_i=1/\lambda_i$. Clearly, $\lambda_i$ correspond to the eigenvalues of $H$. From this, and recalling that the Taylor expansion (\ref{eq:taylor_resolvent}) is centered at zero, we can state that the (\ref{eq:taylor_resolvent}) is valid when the following equivalent conditions hold:
\begin{equation}\label{eq:convergence_condition}
|\mu|<|\mu_i|, \qquad |\lambda|>|\lambda_i| \qquad \qquad \forall i .
\end{equation}
The operator $(I-H)^{-1}$ can be expanded as power series by  noting that $(I-H)^{-1}=-\tilde{R}_1(H)$. Then, by using (\ref{eq:taylor_resolvent}), we recover the Neumann series (\ref{eq:neumann_series}). Since $\lambda=\mu=1$, the convergence condition (\ref{eq:convergence_condition}) becomes $|\lambda_i|<1$, which is equivalent to the classical spectral radius condition for the Neumann series $\rho(H)<1$.

For the sake of compactness, we will now define 
\begin{equation}
v(\mu) \equiv -\tilde{R}_{\mu} c.
\end{equation} 
Clearly, the solution of (\ref{eq:system}) can be obtained as $v(1)$. The approximation of $v(\mu)$ obtained from a $K$ terms truncation of the power series in (\ref{eq:taylor_resolvent}) is written as $v_K(\mu)$. The truncation error of $v_K(\mu)$ is defined as $\epsilon(\mu,K) \equiv \left\|v(\mu)- v_K(\mu) \right\|$ and obeys the following behavior:
\begin{equation}\label{eq:convergence_factor}
\epsilon(\mu,K) = O \left\{ \left( \frac{|\mu|}{\min_i |\mu_i|}  \right)^K  \right\}
\end{equation}
The ratio $|\mu|/|\min_i \mu_i|$ is known as the asymptotic convergence factor \cite[p.~206]{saad_iterative_2003}, or simply convergence factor. In terms of $\lambda$, the convergence factor can be written as $\max_i |\lambda_i|/|\lambda|$. As can be seen from (\ref{eq:convergence_factor}), a smaller convergence factor entails faster convergence.

\subsubsection{Series acceleration}

In this section, we will describe a technique for accelerating the convergence of the power series expansion of the resolvent (\ref{eq:taylor_resolvent}), or even achieving convergence in cases that do not fulfill $\rho(H)<1$. The price to pay for such an improvement is that a certain knowledge of the spectrum of $H$ is required. 

Let us consider a conformal map $t \colon U \to V$ that fulfills $t(0)=0$. With this map, we can define a variable transformation $\tilde{\mu} = t(\mu)$. Since $t$ is a conformal map, there exists an inverse transformation $\mu=t^{-1}(\tilde{\mu})$. Then, $v(\mu)$ can be written as
\begin{equation}\label{eq:tilde_series}
v(\mu) = v(t^{-1}(\tilde{\mu})) = \mu \sum_{k=0}^{\infty} (t^{-1}(\tilde{\mu}) H)^k c. 
\end{equation}
Next, we will find the expansion of $v$ in terms of $\tilde{\mu}$, instead of $\mu$. This can be achieved by directly computing the Taylor series of $v(t^{-1}(\tilde{\mu}))$ in terms of $\tilde{\mu}$ or by suitably rearranging the power expansion in (\ref{eq:tilde_series}). We will use the second procedure. 

Since $t^{-1}$ is a conformal map, it is analytic on its domain. Then, $(t^{-1}(\tilde{\mu}))^k$ can be expanded as a Taylor series:
\begin{equation}\label{eq:tinv_expansion}
(t^{-1}(\tilde{\mu}))^k = \sum_{l=0}^{\infty} \frac{1}{l!} \left. \frac{\partial^l (t^{-1}(\tilde{\mu}))^k}{\partial \tilde{\mu}^k} \right|_{\tilde{\mu}=0} \tilde{\mu}^l.
\end{equation}
It is possible to show that the terms of the series (\ref{eq:tinv_expansion}) are zero for $l<k$, and nonzero in general for $l\geq k$. For this, we use the Generalized Leibniz Rule \cite{thehee_generalization_2002} to write
\begin{equation}\label{eq:leibniz_rule}
\frac{\partial^l (t^{-1}(\tilde{\mu}))^k}{\partial \tilde{\mu}^l} = \sum_{j_1+j_2+\dots+j_k=l} \binom{l}{j_1,\dots,j_k} \prod_{i=1}^{k} \frac{\partial^{j_i}  t^{-1}(\tilde{\mu})}{\partial \tilde{\mu}^{j_i}},
\end{equation}
where the summation runs over the $k$-tuples of non-negative integers $j_i$ that fulfill the condition $\sum_{i=1}^k j_i=l$. For the case $l<k$, since $j_i$ are non-negative integers, then at least one of the elements $j_i$ should be zero to fulfill $\sum_{i=1}^k j_i =l<k$. This implies that, for $l<k$, the zero order derivative (the function $t^{-1}$ itself) will appear at least once on the product on the right side of (\ref{eq:leibniz_rule}). Recalling that $t^{-1}(0)=0$, we conclude (\ref{eq:leibniz_rule}) will be zero when evaluated at $\tilde{\mu}=0$, due to the presence of one or more zero elements on the product. Thus we only need to consider the $l\geq k$ elements in (\ref{eq:tinv_expansion}). Finally, we write (\ref{eq:tinv_expansion}) as
\begin{equation}\label{eq:tinv_tilde_expansion}
\begin{split}
(t^{-1}(\tilde{\mu}))^k &=\sum_{l=0}^{\infty} \frac{1}{l!} \left. \frac{\partial^l (t^{-1}(\tilde{\mu}))^k}{\partial \tilde{\mu}^k} \right|_{\tilde{\mu}=0} \tilde{\mu}^l =  \sum_{l=k}^{\infty}\frac{1}{l!}\frac{\partial^l (t^{-1}(\tilde{\mu}))^k}{\partial \tilde{\mu}^l}\Big|_{\tilde{\mu}=0}\tilde{\mu}^l\\ 
& = \tilde{\mu}^k \sum_{j=0}^{\infty} \frac{1}{(k+j)!}\frac{\partial^{k+j} (t^{-1}(\tilde{\mu}))^k}{\partial \tilde{\mu}^{k+j}}\Big|_{\tilde{\mu}=0}\tilde{\mu}^j = \tilde{\mu}^k \sum_{j=0}^{\infty}\beta_j^{(k)} \tilde{\mu}^{j},
\end{split}
\end{equation}
where we have introduced the coefficients $\beta_j^{(k)}$ for compactness. Now, we introduce (\ref{eq:tinv_tilde_expansion}) into (\ref{eq:tilde_series}) and reorder to obtain a power series in terms of $\tilde{\mu}$:
\begin{equation}\label{eq:mutilde_series}
v(t^{-1}(\tilde{\mu})) = \mu \sum_{k=0}^{\infty} \tilde{\mu}^k\sum_{j=0}^{\infty}\beta_j^k \tilde{\mu}^j H^k c 
=\mu \sum_{n=0}^{\infty}\quad\Big(\sum_{k=0}^n \beta_{n-k}^k (H^k c)\Big) \tilde{\mu}^n= \mu \sum_{n=0}^{\infty} a_n \tilde{\mu}^n,
\end{equation}
where the vector coefficients $a_n$ of the Taylor series in $\tilde{\mu}$ are
\begin{equation}\label{eq:an_coefs}
a_n = \sum_{k=0}^{n} \beta_{n-k}^k H^k c.
\end{equation}

Our ultimate purpose is to compute the solution $u$ of (\ref{eq:system}) as $v(1)$. In the case of the transformed series (\ref{eq:mutilde_series}), $v(t^{-1}(\tilde{\mu}))$ should be evaluated at $\tilde{\mu}=t(1)$. The objective of the transformation in (\ref{eq:taylor_resolvent}) is to improve the convergence of the series: for a suitably chosen $t$, (\ref{eq:mutilde_series}) enjoys better convergence than (\ref{eq:taylor_resolvent}). In particular, we interested in $t$ to provide good converge in the point of interest $\tilde{\mu}=t(1)$.

The convergence conditions of (\ref{eq:mutilde_series}) should be studied in terms of the new variable $\tilde{\mu}$. We can distinguish two cases depending on whether the domain $U$ of the function $t$ contains the singularities $\mu_i$ or not. In both cases, we will assume that the point of interest $\mu$ where we are evaluating $v(\mu)$ belongs to $U$.  

In the first case, we assume that $\mu_i \in U$, so that the images of $\mu_i$ belong to the image of $U$ (Figure tal). Denoting $\tilde{\mu}_i = t(\mu_i)$, we have $\tilde{\mu}_i \in V$. The points $\tilde{\mu}_i$ constitute the singularities of the function $v(t^{-1}(\tilde{\mu}))$. By a similar argument as the used in (\ref{eq:convergence_condition}), the convergence condition for the series (\ref{eq:mutilde_series}) is  
\begin{equation}\label{eq:convergence_condition_tilde}
|\tilde{\mu}|<|\tilde{\mu}_i|, \qquad |\tilde{\lambda}|>|\tilde{\lambda}_i| \qquad \qquad \forall i ,
\end{equation}
where $\tilde{\lambda}$ is defined as $\tilde{\lambda} \equiv 1/\tilde{\mu}$, and $\tilde{\lambda}_i \equiv 1/ \tilde{\mu}_i$. Similarly, the truncation error is
\begin{equation}\label{eq:convergence_factor_tilde}
\epsilon(\tilde{\mu},K) \equiv \left\|  v(t^{-1}(\tilde{\mu})) - v_K(t^{-1}(\tilde{\mu}))  \right\| = O \left\{ \left( \frac{|\tilde{\mu}|}{\min_i |\tilde{\mu}_i|}  \right)^K  \right\}.
\end{equation}

In the second case (Fig tal), we assume the singularities $\mu_i$ do not belong to $U$. Thus, the transformed singularities $\tilde{\mu}_i$ are not defined, and it is not possible to establish the convergence condition (\ref{eq:convergence_condition_tilde}) nor the truncation error (\ref{eq:convergence_factor_tilde}). Instead, it is possible to infer the error from the formalism that will be established in  Section \ref{sec:semiiterative}.  First of all, we write the error of the truncated Neumann series (\ref{eq:truncated_neumann}) as
\begin{equation}\label{eq:iterative_error}
\begin{split}
err_K & \equiv  u_K - u  =  \sum_{k=0}^{K} H^k c - u  =  H\sum_{k=0}^{K-1} H^k c +c - (Hu+c) =\\
& =  H (u_{K-1}-u)  = \dots =  H^K (u_0-u)   =  H^K (c-u) = H^k err_0,   
\end{split}
\end{equation}
where we have defined $err_0\equiv c-u$ as the initial error. For the next step, we will assume that the iteration matrix $H$ can be diagonalized as $H=T^{-1} D T$, where $D$ is a diagonal matrix. Then, we use the semiiterative formalsim (\ref{eq:semiiterative}) and (\ref{eq:iterative_error}) to write the norm error of (\ref{eq:mutilde_series}) as
\begin{equation}\label{eq:cool_error}
\begin{split}
\varepsilon & \equiv \left\| v_K(t^{-1}(\tilde{\mu}))-u\right\| = \left\| \sum_{k=0}^{K}\alpha_{k}^{(K)}u_k - u\right\| = \left\| \sum_{k=0}^{K}\alpha_{k}^{(K)}u_k - \sum_{k=0}^{K} \alpha_{k}^{(K)}u\right\| = \\
& = \left\| \sum_{k=0}^{K} \alpha_{k}^{(K)} \left( u_k - u \right) \right\| = \left\| \sum_{k=0}^{K}\alpha_{k}^{(K)} H^k err_0\right\| = \left\| T^{-1} \left(\sum_{k=0}^{K}\alpha_{k}^{(K)} D^k \right)  T err_0\right\|.
\end{split}
\end{equation}
According to basic linear algebra theory, the elements of the diagonal matrix $D$ are the eigenvalues of $H$. The, taking (\ref{eq:cool_error}) into account, the error $\varepsilon$ will obey
\begin{equation}
\epsilon = O  \left\{\max_i \left| \sum_{k=0}^{K} \alpha_{k}^{(K)} \lambda_i \right| \right\}.
\end{equation}
These expression allows to infer the error decay when the expression (\ref{eq:convergence_factor_tilde}) cannot be applied due to limitations on the domain of $t$.

\subsubsection{Series rearrangement}\label{sec:series_rearrangement}

Up to this point, we have found the solution of (\ref{eq:system}) by using the resolvent formalism, and written the resulting expression as Taylor expansion on an auxiliary variable $\mu$. Then, we have applied a variable transformation to improve the convergence of the series, obtaining the expansion (\ref{eq:mutilde_series}) in terms of $\tilde{\mu}$. However, (\ref{eq:mutilde_series}) is not suitable for practical calculations. Computing the vector coefficients $a_n$ implies a high computational cost, as can be seen from (\ref{eq:an_coefs}). Even if the products $H^k c$ are computed beforehand, computing $a_n$ would imply $n N$ summations (recall that $N$ is the vector size). Thus, the total cost of computing $a_0, \dots, a_K$ for an order $K$ truncation of (\ref{eq:mutilde_series}) would be of order $K^2 N$. The solution is to rearrange (\ref{eq:mutilde_series}) to obtain a series of powers of $H$. So, we rearrange a truncation of (\ref{eq:mutilde_series}) of order $K$:
\begin{equation}\label{eq:gamma_series}
v_K(t^{-1}(\tilde{\mu})) = \mu \sum_{n=0}^{K} a_n \tilde{\mu}^n = \mu \sum_{n=0}^{K} \left(\sum_{k=0}^{n} \beta_{n-k}^k H^k c  \right) \tilde{\mu} = \mu \sum_{k=0}^{K} \left( \sum_{n=k}^{K} \beta_{n-k}^k \tilde{\mu}^n \right) H^k c = \mu \sum_{k=0}^{K} \gamma_k H^k c,
\end{equation}
where the cofficients $\gamma_k$ represent the summation in brackets:
\begin{equation}
\gamma_k^{(K)} \equiv \sum_{n=k}^{K}\beta_{n-k}^{(k)} \tilde{\mu}^{n}
\end{equation}
Note that, whereas the coefficients $a_n$ in (\ref{eq:mutilde_series}) are vectors and involve several products $H^k c$, the coefficients $\gamma_k$ of (\ref{eq:gamma_series}) are scalar. 

\subsubsection{Relation with semiiterative methods}\label{sec:semiiterative}

Next, we will describe the relation between the acceleration method described above and semiiterative methods. This, on one hand, will help us to understand the connection between two families of methods in the literature (series acceleration methods and semiiterative methods) and, on the other, will allow to better study the convergence of the main method of this work. 

Semiiterative methods aim to combine the iterates of an iterative method to obtain a better approximation of the solution \cite[p.~175]{hackbusch_iterative_2016}. In particular, we are interested in linear semiiterative methods, that compute the improved solution as a linear combination of the iterates. Let us assume that a certain iterative method (called primary method) produces the iterates $u_0,\dots,u_K$, that approximate the exact solution $u$. A linear semiiterative method has the form 
\begin{equation}\label{eq:semiiterative}
y_K = \sum_{k=0}^K \alpha_k^{(K)} u_k, \qquad \text{with} \quad \sum_{k=0}^{K} \alpha_k^{(K)} = 1. 
\end{equation} 
It is expected that $y_K$ is a better approximation of the solution $u$ than $u_K$. The condition $\sum \alpha_k^{(K)}=1$ imposes the \textit{constistency} of the method. The semiiterative method (\ref{eq:semiiterative}) can be written as $y_K = F(u_0,\dots,u_K)$. The method is said to be consistent if it fulfills $u=F(u,\dots,u)$. Clearly, a linear semiiterative method fulfills this condition if and only if $\sum \alpha_k^{(K)}=1$ \cite{hackbusch_iterative_2016}. 

The method we have derived in Section \ref{sec:series_rearrangement} can be viewed as a linear semiiterative method. Let us consider the fixed-point iterative method 
\begin{equation}\label{eq:fixed_point}
u_k = H u_{k-1} + c,
\end{equation}
which is equivalent to the Neumann series (\ref{eq:truncated_neumann}). The transformed series (\ref{eq:mutilde_series}) or its rearranged version (\ref{eq:gamma_series}) can be written as a linear combination of the iterates of (\ref{eq:fixed_point}). To see this, let us write the semiiterative method (\ref{eq:semiiterative}) for the primary method (\ref{eq:fixed_point}), and reorder the summation in the following way:
\begin{equation}\label{eq:semiiterative_fixed_point}
y_K = \sum_{k=0}^{K} \alpha_k^{(K)} u_k = \sum_{k=0}^{K} \alpha_k^{(K)} \left( \sum_{j=0}^{k} H^j c  \right) = \sum_{k=0}^{K} \left(  \sum_{j=k}^{K} \alpha_j^{(K)} \right) H^k c. 
\end{equation}
Let us consider the case of interest $\mu=1$. Then, the truncated series (\ref{eq:gamma_series}) can be assimilated to the semiiterative method (\ref{eq:semiiterative_fixed_point}), imposing $v_K(t^{-1}(\tilde{\mu}))=y_K$. By comparing these expressions, we arrive at $\gamma_k^{(K)} = \sum_{j=k}^{K} \alpha_j^{(K)} $. The coefficients $\alpha_k^{(K)}$ in terms of $\gamma_k^{K}$ can be recursively obtained. For $k=K$, $\alpha_K^{(K)}=\gamma_K^{(K)}$. For $k=K-1$, we have $\alpha_{K-1}^{(K)}+\alpha_{K}^{(K)}=\gamma_{K-1}^{(K)}$, which leads $\alpha_{K-1}^{(K)}=\gamma_{K-1}^{(K)}-\gamma_K^{(K)}$. Proceeding recursively, we arrive at
\begin{equation}\label{eq:alphas}
\alpha_k^{(K)}=
\begin{cases}
\gamma_k^{(K)}-\gamma_{k+1}^{(K)}, & \text{if}\ 0\leq k < K \\
\gamma_K^{(K)}, & \text{if}\ k=K 
\end{cases}
\end{equation}
The expression (\ref{eq:alphas}) links the Taylor expansion (\ref{eq:mutilde_series}) with the formalism of semiiterative methods (\ref{eq:semiiterative}): the series (\ref{eq:mutilde_series}) is equivalent to a linear semiiterative method of the form (\ref{eq:semiiterative}), whose $\alpha_k^{(K)}$ coefficients are related to those of (\ref{eq:mutilde_series}) by the expression (\ref{eq:alphas}). Recall that the coefficients $\gamma_k^{(K)}$ are obtained from the conformal map $t\colon U \to V$. 

\subsection{The electromagnetic scattering problem}\label{sec:EM_problem}
So far, we have described a modified Neumann series based on a change of variables by conformal mapping. In this section, we will formulate the electromagnetic scattering problem and will describe how to solve it with the techniques described in Section \ref{sec:iterative_method}.

We formulate the scattering problem with the Volume Integral Equation (VIE) \cite{vanbladel2007electromagnetic}:
\begin{equation}
\frac{1}{j \omega \varepsilon_0(\varepsilon(\mathbf{r})-1)}\mathbf{J}(\mathbf{r})-k^3\int_V \frac{1}{j \omega \varepsilon_0} \overline{\overline{G}}(\mathbf{r},\mathbf{r'}) \cdot \mathbf{J}(\mathbf{r'}) d^3 \mathbf{r'}=\mathbf{E}_{inc}(\mathbf{r}),
\end{equation} 
where $\varepsilon(\mathbf{r})$ stands for the relative permittivity of the scattering body, $\varepsilon_0$ is the vacuum permittivity, $\mathbf{E}_{inc}(\mathbf{r})$ is the electric incident field, $\omega$ is the frequency of the incident field, $k$ is the wavenumber and $\mathbf{J}$ is the electric current (the unknown of the problem). The dyadic Green's function associated with the wave equation is 
\begin{eqnarray}
\overline{\overline{G}}(\mathbf{r},\mathbf{r'}) = (\mathbf{I}+\frac{\nabla \nabla}{k^2}) g(|\mathbf{r}-\mathbf{r'}|)\\
g(|\mathbf{r}|)= \frac{\exp (i k |\mathbf{r}|)}{4 \pi k |\mathbf{r}|}.
\end{eqnarray}
This equation can be written in terms of the electric field as
\begin{equation}\label{eq:integral_equation}
\mathbf{E}(\mathbf{r}) = k^3\int_V(\varepsilon(\mathbf{r})-1) \overline{\overline{G}}(\mathbf{r},\mathbf{r'})\cdot  \mathbf{E}(\mathbf{r'}) d^3 \mathbf{r'}  \mathbf{E}_{inc}(\mathbf{r}). 
\end{equation}
This equation can be written in compact form as
\begin{equation}\label{eq:integral_equation_compact}
\mathbf{E}(\mathbf{r}) = \mathcal{L} \{\mathbf{E}(\mathbf{r}) \} + \mathbf{E}_{inc}(\mathbf{r}), 
\end{equation}
where the linear operator $\mathcal{L}$ has been implicitly defined and stands for the convolution with the dyadic Green's function in \eqref{eq:integral_equation}. This functional equation can be discretized into a linear system with the Method of Moments (MoM) \cite{harrington}. The scattering object is discretized into a mesh of $N$ elements, and \eqref{eq:integral_equation_compact} is projected into a a finite dimensional space formed by a set of test and basis functions associated to the elements of the mesh. In this way, \eqref{eq:integral_equation_compact} is reduced to
\begin{align}\label{eq:integral_equation_discretized}
E &= G E + E_{inc}\nonumber\\
E &= \begin{bmatrix}
\mathbf{E}_{x}\\
\mathbf{E}_{y}\\
\mathbf{E}_{z}
\end{bmatrix}; \qquad \mathbf{E}_{x}, \mathbf{E}_{y}, \mathbf{E}_{z} \in \mathbb{C}^{N}\\
G &= \begin{bmatrix}
\mathbf{G}_{xx} & \mathbf{G}_{xy} & \mathbf{G}_{xz} \\
\mathbf{G}_{yx} & \mathbf{G}_{yy} & \mathbf{G}_{xz} \\
\mathbf{G}_{zx} & \mathbf{G}_{zy} & \mathbf{G}_{zz}
\end{bmatrix} ; \qquad \mathbf{G}_{pq} \in \mathbb{C}^{N \times N}.\nonumber
\end{align}
Although it is not explicitly stated, the incident vector $E_{inc}$ has the same structure as the unknown vector $E$. 

Note that \eqref{eq:integral_equation_discretized} has the form of \eqref{eq:system}. Then, we can apply the iterative method for solving linear systems described in Section \ref{sec:iterative_method} if we know the position of the eigenvalues of $G$. Fortunately, the spectrum of the volume integral operator $\mathcal{L}$ can be found analytically \cite{rahola_eigenvalues_2000}. First of all, let us define as the \textit{constitutive values of the permittivy of an object} as the set of different values that $\varepsilon(\mathbf{r})$ takes in a dielectric object. For instance, for a homogeneous object with permittiviy $\varepsilon=2$, the only constitutive value of its permittivity is 2. 

According to \cite{rahola_eigenvalues_2000}, the eigenvalues of the operator $\mathcal{L}$ for a relatively electrically small object lie in the convex envelope of 0 and the constitutive eigenvalues of its permittivity. For example. For an inhomogeneous  object where the permittivity takes the values $\varepsilon=1+i$ and $\varepsilon=1-i$, the spectrum lies on the triangle with vertices 0, $-1-i$ and $-1+i$. 

The spectral properties of the discretized operator $G$ are similar to those of its continuous equivalent $\mathcal{L}$, so the method described in Section \ref{sec:iterative_method} can be applied to solve \eqref{eq:integral_equation_discretized}. The steps necessary for solving \eqref{eq:integral_equation_discretized} with this method are:
\begin{enumerate}
	\item Compute the convex envelope of 0 and the constitutive values of the permittivity of the object. Let $U$ be the reciprocal image of this convex envelope.
	\item Find a conformal map $t\colon U \to \mathbb{C} \setminus D$ that maps the region $U$ to the complement of the unit circle.
	\item Find, from $t$, the set of coefficients $\gamma_k$.
\end{enumerate}

Finally, the solution of \eqref{eq:integral_equation_discretized} can be computed as 
\begin{equation}\label{eq:EM_gammaseries}
E = \sum_{k=0}^{K} \gamma_k H^k E_{inc}.
\end{equation}

In the particular case where the mesh is regular, then the matrix-vector products $\mathbf{G}_{pq} \mathbf{E}_p$ can be efficiently performed with a Fast Fourier Transform (FFT). If the computational mesh is a cuboid, $\mathbf{G}_{pq}$ has a Toeplitz structure and the operation is straightforward. If it is not a cuboid, it can be trivially enlarged with ghost cells to take advantage of FFT multiplication \cite{chew_fft}. Finally, the matrix-vector multiplication can be efficiently computed as
\begin{equation}\label{eq:multip_fft}
\mathbf{G}_{pq} \mathbf{E}_p = FFT^{-1}(FFT(g_{pq}) \cdot FFT(e_{pq})),
\end{equation}
where $g_{pq}$ and $e_{pq}$ are compressed versions of $\mathbf{G}_{pq}$ and $\mathbf{E}_p$, respectively.

\subsection{Monte Carlo method for electromangetic scattering}\label{sec:EM_MC}

In this section we will describe our Monte Carlo method, and we will succinctly compare it with the classical Ulam-Neumann method. 

Let us consider a multi-dimensional integral 
\begin{equation}\label{eq:deterministic_integral}
\textrm{Int} = \int_{\Omega} f(\mathbf{x}) d\mathbf{x},
\end{equation}
where $\Omega$ is a subset of $\mathbb{R}^d$ and $\mathbf{x} \in \mathbb{R}^d$ \cite{evans_MC}. We approximate  \eqref{eq:deterministic_integral} with the estimator
\begin{equation}
Q_S(f) = \frac{\textrm{Vol}(\Omega)}{S} \sum_{s=0}^{S-1} f (\mathbf{X}_s),
\end{equation} 
where $\mathbf{X}_s$ are independent random samples from $\Omega$ with uniform distribution, $\textrm{Vol}(\Omega) = \int_{\Omega} d \mathbf{x}$ is the volume of the subset $\Omega$, and $S$ is the number of Monte Carlo samples. It is known that, if $f$ fulfills certain regularity conditions, $Q_S(f)$ tends the exact result $\textrm{Int}$ of \eqref{eq:deterministic_integral} as $S \to \infty$. Furthermore, it converges as
\begin{equation}\label{eq:mc_convergence}
|\textrm{Int} - Q_s(f)| = \frac{\sigma(f)}{\sqrt{S}},
\end{equation}
where $\sigma(f)$ is the standard deviation of $f$ \cite{leobacher_introduction_2014}. It is also possible to accelerate the convergence of Monte Carlo integration by using importance sampling. Instead of using a uniformly distributed probability, importance sampling takes probability distributions that concentrate the sampling in regions of $\Omega$ that are more relevant for the integration. In the case of importance sampling, the Monte Carlo approximation of the integral is written as 
\begin{equation}\label{eq:importance_sampling}
Q_S(f) = \frac{1}{S} \sum_{s=0}^{S-1} \frac{f(\mathbf{X}_s)}{p(\mathbf{X}_s)},
\end{equation}
where $p(\mathbf{X}_s)$ is the probability of selecting $\mathbf{X}_s$. 

Our method relies on an importance sampling strategy applied to the different scattering terms of \eqref{eq:EM_gammaseries} . The final objective is not to obtain the electric field itself, but other derived electromagnetic quantities that characterize the system, as the Scattering Cross Section (RCS) \cite{balanis2012advanced}. The RCS, as well as other quantities, can be computed from the radiation vector $\mathbf{N}(\hat{\mathbf{r}})$ defined as 
\begin{equation}\label{eq:radiation_vector}
\mathbf{N}(\hat{\mathbf{r}}) = - i \omega \varepsilon_0 \int_{V} (\varepsilon(\mathbf{r}')-1) \mathbf{E}(\mathbf{r}') \exp(i k \hat{\mathbf{r}} \cdot \mathbf{r}') d \mathbf{r}'. 
\end{equation}
If we consider the discretized version of $\mathbf{E}(\mathbf{r})$, the integral \eqref{eq:radiation_vector} results in a discrete summation
\begin{equation}\label{eq:radiation_vector_discrete}
\mathbf{N}(\hat{\mathbf{r}}) = \sum_j w_j E_j,  
\end{equation}
where $w(j) = -i \omega \varepsilon_0 (\varepsilon(\mathbf{r}_j)-1) \exp(i k \hat{\mathbf{r}} \cdot \mathbf{r}_j)$. Equivalently, this can be written as a scalara product $w \cdot E$, where $w$ is the vector of $w_i$ elements. If we use \eqref{eq:EM_gammaseries} to write the electric field and insert it in \eqref{eq:radiation_vector_discrete} we obtain
\begin{equation}\label{eq:radiation_vector_expanded}
\mathbf{N}(\hat{\mathbf{r}}) =  \big(\sum_{k=0}^{K} w \cdot G^k E_{inc} \big). 
\end{equation}
Each one of the terms of the series \eqref{eq:radiation_vector_expanded} can be thought as a multiple dimensional integral. Thus the terms can be approximated by Monte Carlo by using \eqref{eq:importance_sampling}. This sampling takes into account the tensorial nature of $G$ (it is divided in $\mathbf{G}_{pq}$ blocks). Let $\xi_s$ be a Markov of fixed length $k$: $\xi = \xi^0 \to \xi^1 \to \dots \to \xi^k$ where $\xi^l$ are integer numbers between 1 and $N$ (the number of elements on the mesh)  and let $p(\xi)$ the probability of obtaining a particular Markov chain. Also, let us define the sample weight
\begin{equation}
T(\xi) = w_{\xi^0}\Bigg(\prod_{l=1}^{k}\begin{bmatrix}
\mathbf{G}_{xx}(\xi^{l-1},\xi^l) & \mathbf{G}_{xy}(\xi^{l-1},\xi^l) & \mathbf{G}_{xz}(\xi^{l-1},\xi^l) \\
\mathbf{G}_{yx}(\xi^{l-1},\xi^l) & \mathbf{G}_{yy}(\xi^{l-1},\xi^l) & \mathbf{G}_{xz}(\xi^{l-1},\xi^l) \\
\mathbf{G}_{zx}(\xi^{l-1},\xi^l) & \mathbf{G}_{zy}(\xi^{l-1},\xi^l) & \mathbf{G}_{zz}(\xi^{l-1},\xi^l)
\end{bmatrix}\Bigg)  \begin{bmatrix}
\mathbf{E}_{inc,x}(\xi^k)\\
\mathbf{E}_{inc,y}(\xi^k)\\
\mathbf{E}_{inc,z}(\xi^k)
\end{bmatrix}
\end{equation}
for $k\geq1$. For the case $k=0$, $T(\xi)$ reduces to
\begin{equation}
T(\xi) = w_{\xi^0}\begin{bmatrix}
\mathbf{E}_{inc,x}(\xi_0)\\
\mathbf{E}_{inc,y}(\xi_0)\\
\mathbf{E}_{inc,z}(\xi_0)
\end{bmatrix}.
\end{equation}
Note that the sample weight $T(\xi_s)$ is a vector of length $3$, since the sample of the incident field has length $3$ and the sample of the $G$ matrix is $3  \times 3$. With this, we define the estimator of $w \cdot G^k E_{int}$ based on $S$ independent Markov chains $\xi_s$
\begin{equation}
w \cdot (G^k E_{int}) \approx  \frac{1}{S} \sum_{s=0}^{S-1} \frac{1}{p(\xi_s)} T(\xi_s).
\end{equation} 
The probability  $p(\xi)$ can be computed as $p(\xi) = p(\xi^0)p(\xi^0,\xi^1)p(\xi^1,\xi^2)\dots p(\xi^{k-1},\xi^k)$. Here, $p(\xi^0)$ represents the probability of $\xi^0$ to be the first state of the Markov chain. We will assume that this is a uniformly distributed random variable. Since $\xi^0$ can take $N$ different values, $p(\xi_0) = 1/N$. $p(\xi^{l-1},\xi^{l})$ represents the probability transition of moving to a state $\xi^l$ from a given state $\xi^{l-1}$. The transition probabilities are usually represented by a transition matrix
\begin{equation}
P = \begin{bmatrix}
p(1,1) & \dots & p(1,N)\\
\vdots  & \ddots  & \vdots \\
p(N,1) & \dots & p(N,N)
\end{bmatrix} 
\end{equation}
that fulfills the condition $\sum_j p(i,j) = 1,  \forall i$. The matrix $P$ determines the importance sampling strategy, since it establishes which elements of the series \eqref{eq:EM_gammaseries} are more likely to be sampled. The elements of $G$ with larger absolute value (corresponding to stronger electromagnetic interactions) should be sampled more frequently than those with smaller absolute value. For this, we employ Monte Carlo Almost Optimal (MAO) transition matrix \cite{dimov_book}. With this technique, the elements of the probability transition matrix are
\begin{equation}
p(i,j) = \frac{\Bigg\| 
	\begin{bmatrix}
	\mathbf{G}_{xx}(i,j) & \mathbf{G}_{xy}(i,j) & \mathbf{G}_{xz}(i,j) \\
	\mathbf{G}_{yx}(i,j) & \mathbf{G}_{yy}(i,j) & \mathbf{G}_{xz}(i,j) \\
	\mathbf{G}_{zx}(i,j) & \mathbf{G}_{zy}(i,j) & \mathbf{G}_{zz}(i,j)
	\end{bmatrix}
	\Bigg\|}{ \sum_{k=1}^{N} \Bigg\| 
	\begin{bmatrix}
	\mathbf{G}_{xx}(i,k) & \mathbf{G}_{xy}(i,k) & \mathbf{G}_{xz}(i,k) \\
	\mathbf{G}_{yx}(i,k) & \mathbf{G}_{yy}(i,k) & \mathbf{G}_{xz}(i,k) \\
	\mathbf{G}_{zx}(i,k) & \mathbf{G}_{zy}(i,k) & \mathbf{G}_{zz}(i,k)
	\end{bmatrix}
	\Bigg\|},
\end{equation} 
where $\|\bullet\|$ stands for the matrix norm. Since $P$ is a dense matrix, the cost of exactly computing the matrix $P$ is prohibitively large. So, compression strategies are used. 

Note that the convergence condition for the Ulam-Neumann method is $\rho(G^{*}<1)$, where $G^*_{ij}=G^2(i,j)/p(i,j)$ \cite{ji_convergence_2013}, whereas our independent estimations convergence if the integrated function is $L^2$ integrable.   

\section{Numerical experiments}\label{sec:numerical_experiments}

In this section we present numerical results for the method \eqref{eq:EM_gammaseries} both for its deterministic and Monte Carlo implementations. The discretizations of the scattering operator are carried out by the Galerkin Method \cite{rahola_eigenvalues_2000,sertel_integral_2012}. 

\subsection{The modified Born series}

\begin{figure}[h]
	\centering\includegraphics[height=8.5cm]{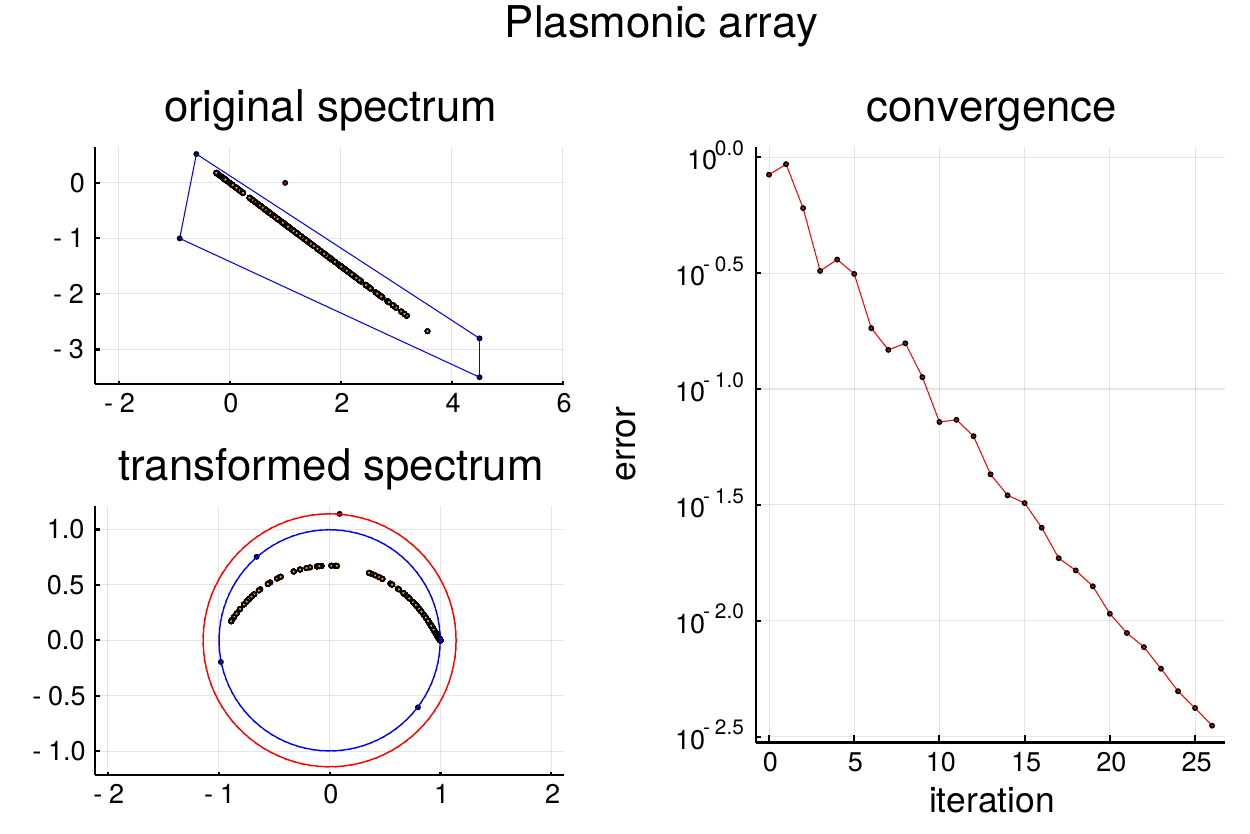}
	\caption{Plasmonic array. Upper left: original spectrum of the system (points) and polygon used for the conformal transformation (blue lines). Note that the eigenvalues lie on a single line as predicted by the spectral localization theorem. Down left: singularities after the transformations (points), limit for convergence (red line) and transformed polygon (blue lines). Right: convergence of \eqref{eq:EM_gammaseries} up to order 26.}
	\label{fig:plasmonic_array}
\end{figure}

Before testing the Monte Carlo algorithm, we test the modified Born series. Our first example is a homogeneous array of square patches with plasmonic permittivity ($\varepsilon=-3.0+3.0i$). We consider 100 patches distributed in a $10\times 10$ array. Each patch has a side length of $10$ nm and a thickness of 1 nm. The patches are distribute in a square array with a lattice constant of 10 nm. They are illuminated from the top with a plane wave with wavelength $\lambda = 400$ nm. The conformal transformation is build from a polygon of four points: $4.5-2.8i$, $-0.6+0.52i$, $-0.9-1.0i$, $4.5-3.5i$. This polygon encloses the eigenvalues and the transformation maps them inside the unit circle (Figure \ref{fig:plasmonic_array}). The transformation results in $\tilde{\mu}=0.0690-0.8716i$. The convergence of \eqref{eq:EM_gammaseries} for this transformation is also displayed in Figure  $\ref{fig:plasmonic_array}$. The $\gamma_k$ coefficients associated to this transformation up to order 9 are displayed in Table \ref{tab:gamma_table}.

\begin{figure}[h]
	\centering\includegraphics[height=8.5cm]{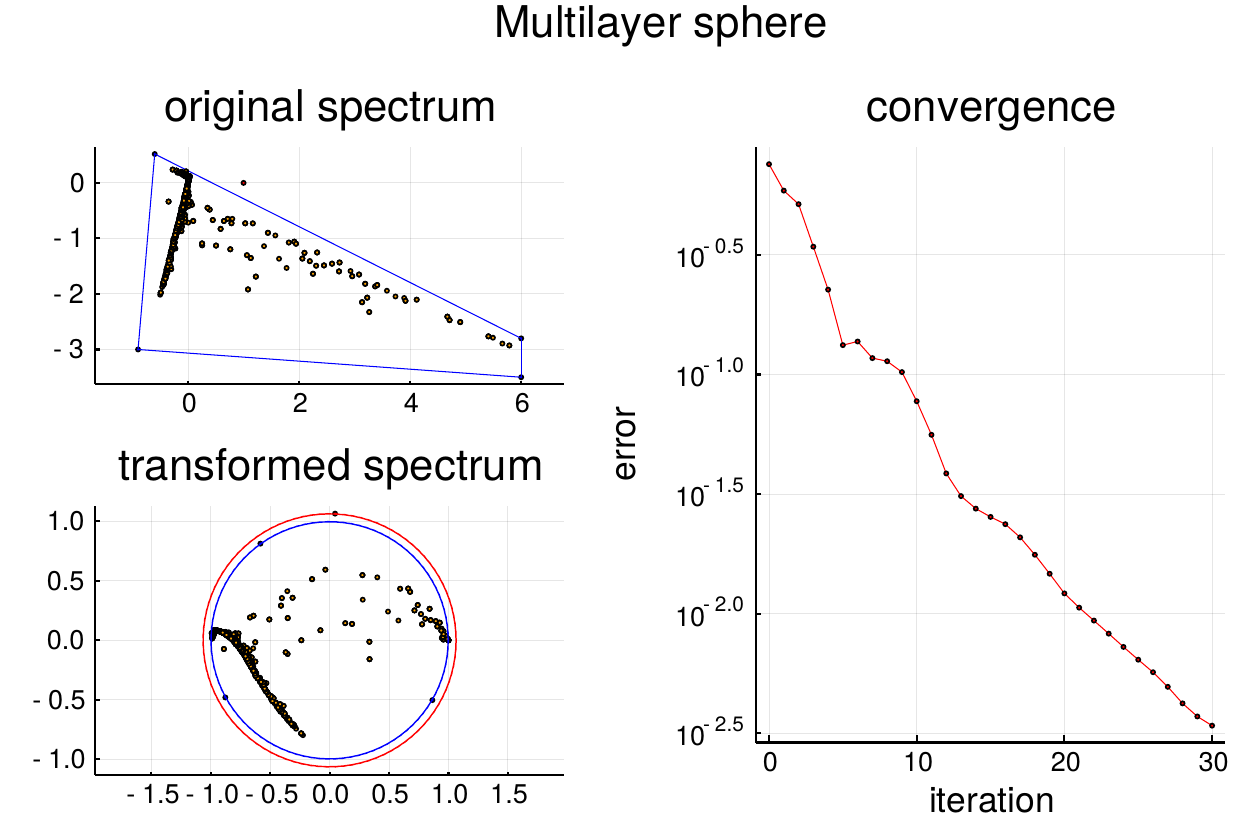}
	\caption{Multilayer sphere. Upper left: original spectrum of the system (points) and polygon used for the conformal transformation (blue lines). Note that the eigenvalues  lie withing the convex envelop of zero and the constitutive values of the permittivity with a minus sign, as predicted by the spectral localization theorem. Down left: singularities after the transformations (points), limit for convergence (red line) and transformed polygon (blue lines). Right: convergence of \eqref{eq:EM_gammaseries} up to order 26.}
	\label{fig:multilayer_sphere}
\end{figure}

Our second case corresponds to a double-layered inhomogeneous sphere. The inner core of the sphere has a radius of 30 nm and a permittivity of $\varepsilon=-5.0+3.0i$ (plasmonic). The outer shell has an external radius of $60$ nm and a permittivity of $\varepsilon=1.5+2.0i$. The sphere is discretized into a regular computational mesh of $30 \times 30  \times 30$. It is illuminated with a plane wave of $\lambda=400$ nm. The conformal transformation is build from a polygon of four points: $6.0-2.8i$, $-0.6+0.52i$, $-0.9-3.0i$, $6.0-3.5i$. This polygon encloses the eigenvalues and the transformation maps them inside the unit circle (Figure \ref{fig:plasmonic_array}). The transformation results in $\tilde{\mu}=0.0403-0.9366i$. The convergence of \eqref{eq:EM_gammaseries} for this transformation is also displayed in Figure  $\ref{fig:multilayer_sphere}$. The $\gamma_k$ coefficients associated to this transformation up to order 9 are displayed in Table \ref{tab:gamma_table}.

\begin{table}[h!]
	\begin{center}
		\caption{Values for the $\gamma_k$ for the plasmonic array and the inhomogeneous case.}
		\label{tab:gamma_table}
		\label{tab:table2}
		
		\begin{tabular}{|c|c|c|} 
			& Plasmonic array & Multilayer sphere  \\
			\hline
			$\gamma_0$ & $ 1.0+0.0i$ & $1.0+0.0i$ \\
			$\gamma_1$ & $ 1.1807+0.0024i$ & $ 1.2858-0.1280i$ \\
			$\gamma_2$ & $ 1.5786-0.6495i$ & $ 1.3104-1.1235i$  \\
			$\gamma_3$ & $ 0.7290-1.9407i$ & $ -0.2831-1.9918i$  \\
			$\gamma_4$ & $ -1.4212-1.4946i$ & $ -1.7673+0.0600i$  \\
			$\gamma_5$ & $ -1.2755+0.9392i$ & $ 0.2288+0.9713i$  \\
			$\gamma_6$ & $ 0.6111+0.5928i$ & $  0.2962-0.2042i$  \\
			$\gamma_7$ & $ 0.1060+0.2734i$ & $ -0.0741-0.0384$  \\
			$\gamma_8$ & $-0.0586+0.0115i$ & $0.0013+0.0120i$  \\
			$\gamma_9$ & $+0.0049+0.0042i$ & $0.0007-0.0006$  \\
			\hline
		\end{tabular}
	\end{center}
\end{table}

\subsection{The Monte Carlo Algorithm}

\begin{figure}[h]
	\centering\includegraphics[height=8.5cm]{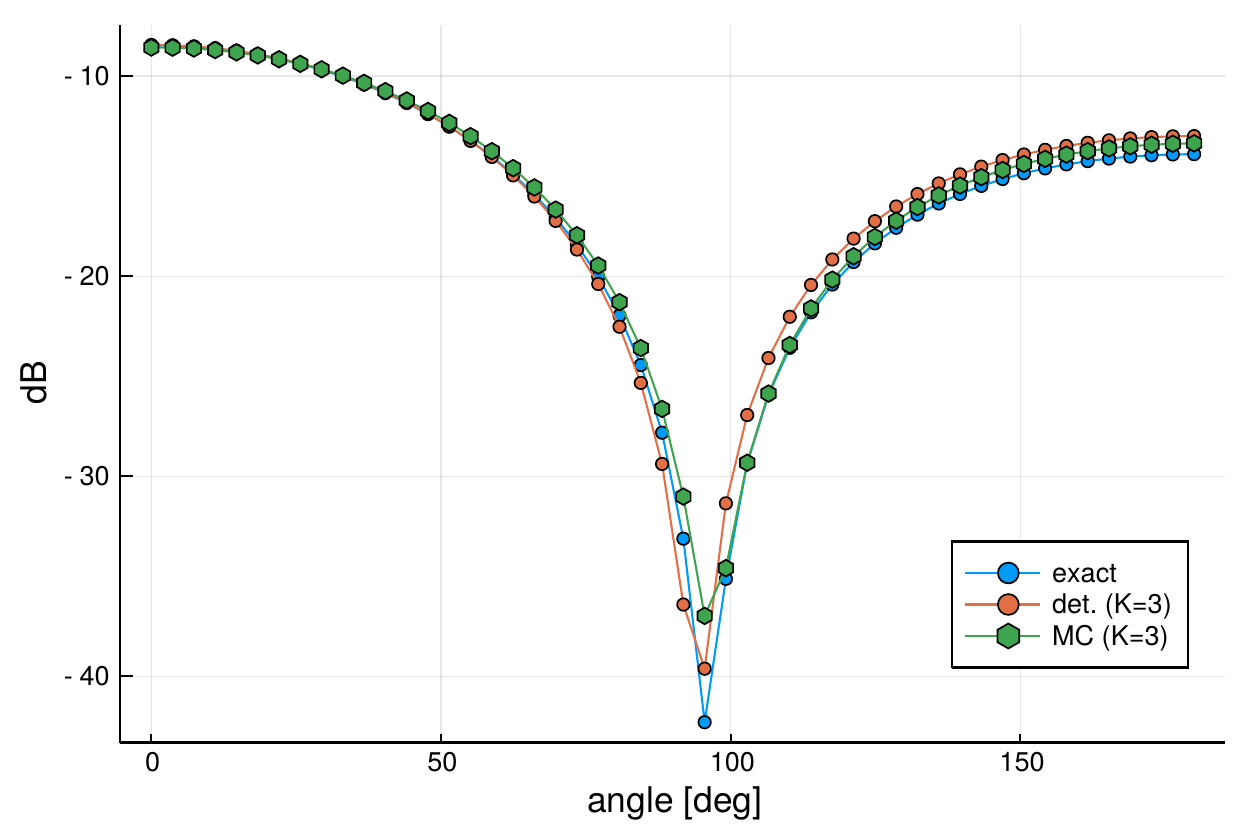}
	\caption{BiStatic Radar Cross Section for a dielectric cube of $\varepsilon=4.0$ and side $\lambda/4$ \cite{zwamborn}. The exact result, the result obtained by \eqref{eq:EM_gammaseries}, and the result obtained by Monte Carlo integration are displayed. All the three results are practically equivalent: the relative error between the exact result and the Monte Carlo result is 0.039.}
	\label{fig:MC_cube}
\end{figure}

We will apply the Monte Carlo described in Section \ref{sec:MC_method} to a dielectric cube of permittivity $\varepsilon=4.0$ and side length $\lambda/4$, impinged by a plane wave. This object is used to test electromagnetic scattering algorithm in works as \cite{chew_fft,zwamborn}. The objective is to compute the BiRCS of this object. The object is discretized into a regular computational mesh of $32\times 32 \times 32$. The polygon associated with the conformal mapping transformation has the vertices $0.5+0.9i$, $-3.5+0.9i$, $-3.5-0.8i$, $0.5-0.8i$. We set $K=3$ and the number of Monte Carlo samples is $ S = 15 \cdot 10^6$. 

The results are shown in Figure \ref{fig:MC_cube}. For comparing purposes, we plot the exact RCS (computed with a FFT accelerated GMRES with error tolerance $10^{-16}$), the BiRCS computed with the deterministic formula \eqref{eq:EM_gammaseries}, and the BiRCS computed with the Monte Carlo method. The relative error between the Monte Carlo result and the exact result is 0.039, and the relative error between the result computed with \eqref{eq:EM_gammaseries} and the exact result is $0.067$. 

Regarding the computational cost, the deterministic method \eqref{eq:EM_gammaseries} applied with FFT acceleration is more efficient than its equivalent Monte Carlo method when executed serially, by a factor 1.72. If the method \eqref{eq:EM_gammaseries} is implemented without FFT acceleration, then the Monte Carlo method is much more efficient, and represents only a $0.15\%$ of the computations required by the deterministic method.  

Also, we have experimentally observed that the error of the Monte Carlo method decays as $\sqrt{S}$, according to \eqref{eq:mc_convergence}

\begin{figure}[h]
	\centering\includegraphics[height=8.5cm]{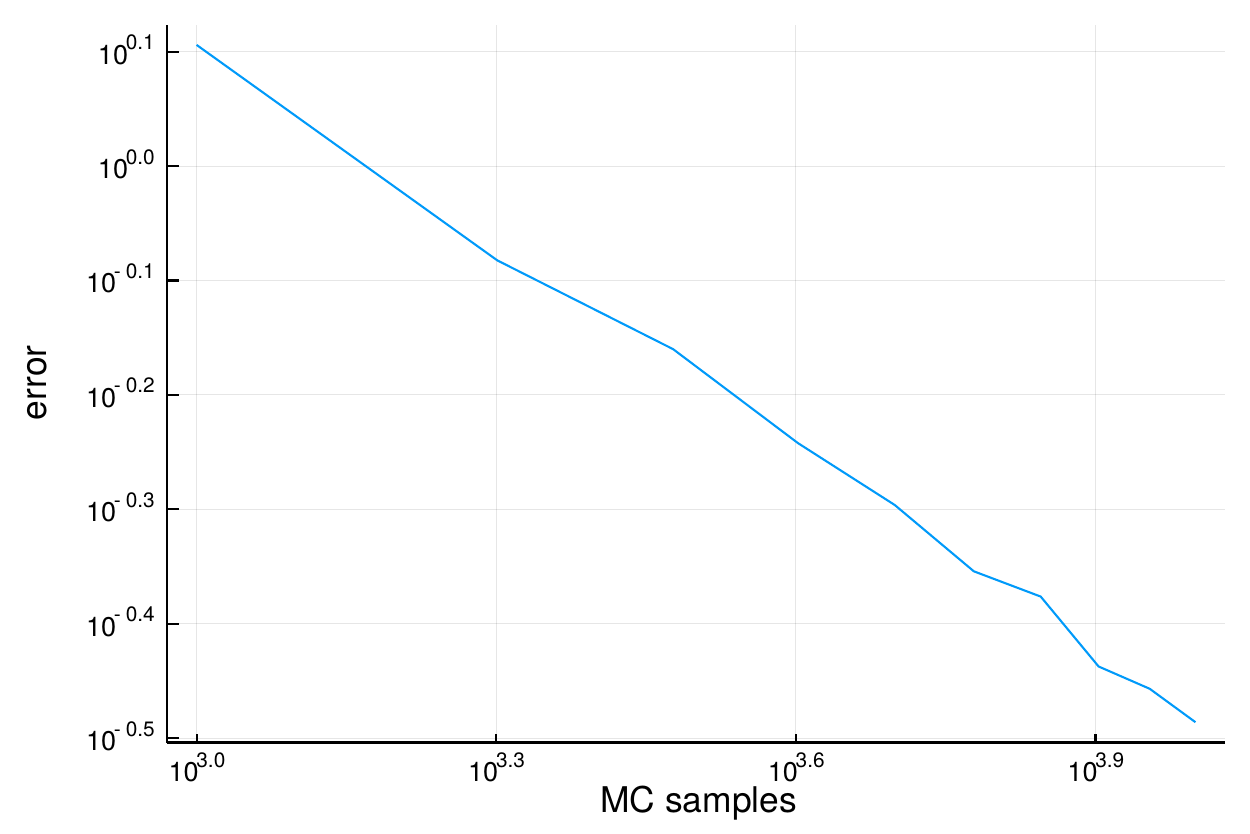}
	\caption{Convergence of the Monte Carlo algorithm for the dielectric cube of $\varepsilon=4.0$ in a computational mesh $10 \times 10 \times 10$. The error of Monte Carlo integration decays as $\sqrt{S}$}
	\label{fig:MC_convergence}
\end{figure}

\newpage

\section{Scalability}\label{sec:scalability}

In this section we derive an analytical performance model that describes how the Monte Carlo algorithm and its corresponding exact deterministic counterpart would scale in a parallel multicore environment. The model accounts for both computation and communication costs. 

\subsection{Deterministic method}\label{sec:deter_method}

The exact deterministic method corresponds to equation  \eqref{eq:EM_gammaseries}. If the cells of the discretization grid form a regular lattice, the matrix-vector products between $\mathbf{H}_{pq}$ and $\mathbf{E}_p$ can be efficiently computed with the help of the Fast Fourier transform \cite{chew_fft}. Our model is based on the performance model for 3D FFTs of \cite{Czechowski}, but accounts for the particular features of the algorithm under study. 

Let us assume that the discretization grid consists on $n \times n \times n$ points, and that the computational environment has $P$ nodes or computing units. The 3D FFT can be computed by performing three sets of 1D FFTs (computation phases), separated by two communication phases. 

Each computation phase comprises $n^2$ 1D FFTs distributed among $P$ nodes. A 1D FFT computed with the radix-2 or Cooley-Tukey algorithm has an approximate cost of $5n \log_2 n $ operations. Then, according to \cite{Czechowski}, the total computation time associated to a 3D FFT is
\begin{equation}
T_{\textrm{3DFFT}}^{\textrm{calc}} = 3 \cdot \frac{n^2}{P} \cdot \frac{5n\log n}{C_{\textrm{node}}},
\end{equation}    
where $C_{\textrm{node}}$ is the node performance in FLOPS and the $3$ factor accounts for the $3$ computation phases. 

In the communication phase, each node performs a personalized all-to-all exchange of its data with $\sqrt{P}$ other nodes. For modeling purposes, the amount of bandwidth available in the network can be approximated by the bisection bandwidth \cite{Solihin,Grama}. Thus, the effective bandwidth $\beta$ is computed as $\beta = \beta_{\textrm{bw}} \beta_{\textrm{link}}$, where $\beta_{\textrm{bw}}$ is the bisection bandwidth and $\beta_{\textrm{link}}$ is the bandwidth of the link. The bisection bandwidth can be analytically computed and depends on the network topology and the number of nodes. For instance, a $d$-dimensional torus with $P$ nodes in total has a bisection bandwidth $\beta_{\textrm{bw}} = 2 p ^{\frac{d-1}{d}}$ \cite{Grama}. The total time spent in the two communication phases is
\begin{equation}
T_{\textrm{3DFFT}}^{\textrm{comm}} = 2 \cdot \frac{n^3}{\beta(P)}.
\end{equation}
Finally, the total cost of the 3D FFT is
\begin{equation}
T_{\textrm{FFT}} = T_{\textrm{FFT}}^{\textrm{calc}}+T_{\textrm{FFT}}^{\textrm{comm}}. 
\end{equation}

The iterative method \eqref{eq:EM_gammaseries} requires several FFTs, inverse FFTs and additional communication operations. In the set up phase, nine 3D FFTs associated to the nine submatrices $\mathbf{H}_{pq}$ would be performed. The Fourier compressed versions of the $\mathbf{H}_{pq}$ blocks are used at each iteration step $k$, but only need to be computed once. Then, the total cost of the setup phase is $9 T_{\textrm{3DFFT}}$. 

Each iterative step comprises nine matrix-vector products between the $\mathbf{H}_{pq}$ blocks and the $\mathbf{u}_{k,p}$ vectors, that will be efficiently computed with \eqref{eq:multip_fft}. The transformed elements $FFT(h_{pq})$ had been computed in the setup phase, but it is necessary to perform three 3D FFTs to obtain the transformed vectors $FFT(u_{k,p})$ for $p=x,y,z$. Besides, following \eqref{eq:multip_fft}, nine inverse 3D FFTs are needed to obtain the nine matrix-vectors products. Assuming that both the 3D FFT and its inverse transform have the same cost, the transforms and antitransforms at each $k$ step cost $3T_{\textrm{FFT}}+9T_{\textrm{FFT}}=12 T_{\textrm{FFT}}$. Furthermore, we also need to take into account the computation and communication costs for adding the nine vectors of $n^3$ elements resulting from the matrix-vector products, it is, the operations $\sum_{i=x,y,z} \mathbf{H}_{pi} \mathbf{u}_i$ for $p=x,y,z$. Each one of the  $\mathbf{H}_{pi} \mathbf{u}_i$ vectors has a total of $n^3$ elements, so the summation of three components ($i=x,y,z$) has a cost of $2n^3$ operations.  Since three summations are performed ($p=x,y,z$), the total number of operations of this phase is $6n^3$. Communication cost strongly depends on implementation. Here we will assume that only a two thirds of the $9 n^3$ available data points need to be sent from one processor to another. The reason is that, in the summation $\sum_{i=x,y,z} \mathbf{H}_{pi} \mathbf{u}_i$, one of the three matrix-vector products will remain in the location where it has been computed, and the other two will be sent to the location of the former one to perform the addition. Then, in this phase it is necessary to send through the network a total of $2/3 \cdot 9 n^3 = 6n^3$ data points. The time spent in the summation phase at each $k$ step are
\begin{equation}
T_{\textrm{add}}^{\textrm{calc}} = 6 \cdot \frac{n^3}{P C_{\textrm{node}}}; \qquad T_{\textrm{add}}^{\textrm{comm}} = 6 \cdot \frac{n^3}{\beta(P)}.
\end{equation}

Finally, the total parallel time of the deterministic method \eqref{eq:EM_gammaseries} is the sum of the setup phase time and the time corresponding to $K$ iterative steps:
\begin{equation}\label{eq:time_det}
T_{\textrm{det}}(P) = 9 T_{\textrm{3DFFT}} + K\cdot(12 T_{\textrm{3DFFT}} +  T_{\textrm{add}}^{\textrm{calc}} + T_{\textrm{add}}^{\textrm{comm}})
\end{equation}

\subsection{Monte Carlo method}\label{sec:MC_method}

Due to its simplicity, the performance model of the Monte Carlo method is easier to obtain. As Monte Carlo is an embarrassingly parallel method, there is inter-node communication cost. Yet, some communications will be needed at the end of the execution to gather all the samples for computing the final result (in this case at BiRCS value). However, these are synchronization operations in the terminology of \cite{Yavits}, and we neglect them both in the deterministic and the Monte Carlo performance models. 

Let $S$ be the number of random samples, and let $K$ be the order of the sampled series \eqref{eq:EM_gammaseries}. As stated in section (alguna), sampling the zero order term of \eqref{eq:convergence_condition} has a cost of $3$ operations per sample, whereas sampling terms with $k\geq0$ has a cost of $9k$ operations per sample, since each sample consists of a $3\times 3$ matrix. Assuming for simplicity that we take the same number of samples for each order of \eqref{eq:EM_gammaseries}, the total number of operations in the Monte Carlo method is
\begin{equation}
3 S + 9 S + 2 \cdot 9 S + \dots + K 9 S = 3 S \Big(1 + \frac{3(K+1)K}{2}\Big)
\end{equation} 
and the total parallel time of the Monte Carlo method is
\begin{equation}\label{eq:time_MC}
T_{\textrm{MC}}(P) = \frac{S}{P C_{\textrm{node}}} 3 \Big(1 + \frac{3(K+1)K}{2}\Big) 
\end{equation}
Note that, from \eqref{eq:time_MC}, the Monte Carlo method shows linear scalability. 
\subsection{Comparison}

Here, we compare the scalability of the deterministic and the Monte Carlo method for a particular case. As in the example of (section alguna) we consider a dielectric cube with $\varepsilon = 4.0 $ and side $\lambda/4$ irradiated with a plane wave. This object is frequently employed as an example to validate numerical methods, as inx \cite{chew_fft}. The object is discretized into a mesh of $32 \times 32 \times 32$ cells. In the case of the Monte Carlo method, we take $S = 5 \times 10^{6}$. 

Regarding the parameters $\beta_{\textrm{link}}$ and $C_{\textrm{node}}$, we take the values from \cite{Yavits} $\beta_{\textrm{link}} = 21.3 GB/s$ and $C_{\textrm{node}}=50.4 GF/s$. The work \cite{Czechowski}, which studies the potential scalability of 3D FFT in a hypothetic exascale machine, dates from 2012, so the numerical values of the hardware parameters may be outdated. However, using the values proposed in \cite{Czechowski} has an advantage: these values have been adjusted to their effective value associated to this particular application. We are interested in scalability rather than in absolute computation time, so our model, in combination with the parameter values in \cite{Czechowski} should suffice to draw some conclusions about the behavior of our algorithm.

Also, the example under consideration here is not large enough to be executed in a machine with thousands of cores. But again, we are more interested in the asymptotic behavior of the model rather than in absolute numbers. We choose this particular example for being a representative problem in this field.

\begin{figure}[h]
	\centering\includegraphics[height=8.5cm]{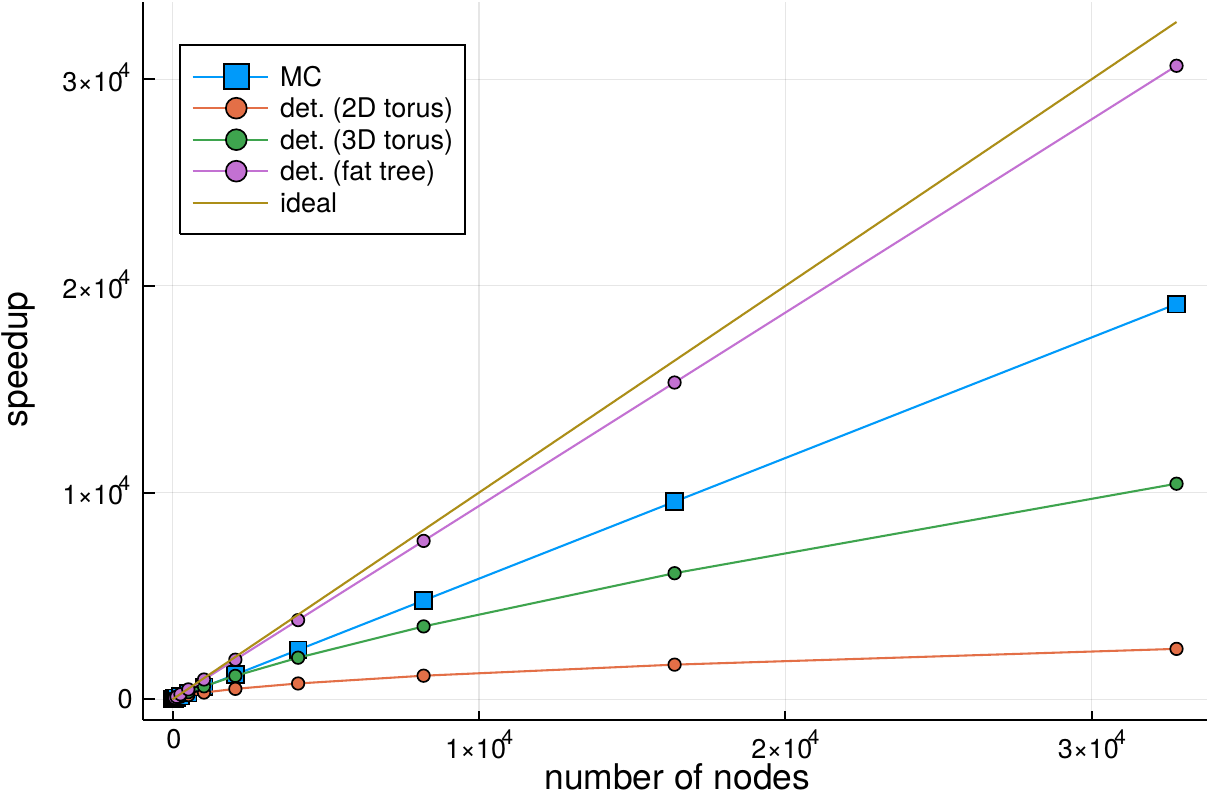}
	\caption{Predicted scalability of the Monte Carlo method (MC) and the deterministic FFT-accelerated method (det.) for various network topologies, according to \eqref{eq:speedup}. The MC method outperforms the deterministic method in topologies with with worse bisection bandwidth (2D and 3D torus). In the case of a fat tree topology the deterministic method performs better than Monte Carlo, but at the cost of a more complex network.}
	\label{fig:scalability}
\end{figure}

Figure \ref{fig:scalability} shows the scalability of the deterministic method (Section \ref{sec:deter_method}) and the Monte Carlo method (Section \ref{sec:MC_method}). The speedup of the deterministic method is computed for several network topologies, while the performance of the Monte Carlo method is independent of the network, as there are no inter-node communications. In order to properly compare the performance of both methods, the speedup is computed with respect to the serial time for \eqref{eq:time_det} for both the deterministic and Monte Carlo algorithms:
\begin{equation}\label{eq:speedup}
\textrm{speedup det. } = \frac{T_{\textrm{det}}(1)}{T_{\textrm{det}}(P)}; \qquad \textrm{speedup MC} = \frac{T_{\textrm{det}}(1)}{T_{\textrm{MC}}(P)}
\end{equation}
As the number of nodes increases, the Monte Carlo method shows a better behavior than the deterministic method for topologies whose bisection bandwidth $\beta_{\textrm{bw}}$ scales sublinearly with $P$, as toroidal topologies. In a fat tree topology ($\beta_{\textrm{bw}}=P/2$) the deterministic method outperforms the Monte Carlo method. However, both the Monte Carlo method and the fat tree deterministic method scale linearly with the number of nodes. This implies that the Monte Carlo method has a better compromise performance and network usage. 

Besides, there are many effects that have not been taken into consideration and can degrade the performance of the deterministic method, making the Monte Carlo one relatively better. For example, \cite{Czechowski}, which is the base of our performance model, claims that the computation phase in a 3D torus topology incurs in a large overhead that is not explained by the analytical model. Also, \cite{ayala} shows how the performance of a 3D FFT application (FFTMPI) drops for a few hundreds of nodes in Summit, a machine with fat tree topology, due to latency effects. Due to its embarrassingly parallel nature and its ease for synchronization, the Monte Carlo method is not likely to be affected by network limitations. Besides, it is well suited for fine granularity systems as GPUs.          

\section{Conclusions}\label{sec:conclusions}

In this paper we have proposed  a Monte Carlo method for solving the electromagnetic scattering problems for dielectric objects. Our method relies on a modified Born series based on a conformal mapping transformation. Thanks to the spectral localization theorem for the Volume Integral Equation (VIE)formulation, it is possible to compute beforehand the coefficients associated to the modified Born series. This allows us to approximate each one of the terms of the series --corresponding to a different order of scattering-- with Monte Carlo methods. Numerical examples show the validity the algorithm. 

The main advantages of this method are its ease of formulation and its potential parallelism. The Monte Carlo samples can be independently computed without the need of inter-processor communications. We have derived an analytic scalability model that shows how the absence of communications will allow this method to scale better than other algorithms that are more efficient when executed serially, but suffer communication overhead at large scale execution.

\bibliography{bibliofile.bib}

\providecommand{\newblock}{}
\begin{thebibliography}{10}
\expandafter\ifx\csname url\endcsname\relax
  \def\url#1{{\tt #1}}\fi
\expandafter\ifx\csname urlprefix\endcsname\relax\def\urlprefix{URL }\fi
\providecommand{\eprint}[2][]{\url{#2}}

\bibitem{zettili_quantum_2009}
Zettili N 2009 {\em Quantum mechanics: concepts and applications\/} 2nd ed
  (Chichester, U.K: Wiley) ISBN 978-0-470-02678-6 978-0-470-02679-3 oCLC:
  ocn255894625

\bibitem{mishchenko_libro_gordo}
Mishchenko M~I 2014 {\em Electromagnetic Scattering by Particles and Particle
  Groups: An Introduction\/} (Cambridge University Press) ISBN 9780521519922

\bibitem{Kouri}
Kouri D~J and Vijay A 2003 {\em Phys. Rev. E\/} {\bf 67}(4) 046614
  \urlprefix\url{https://link.aps.org/doi/10.1103/PhysRevE.67.046614}

\bibitem{hutson_applications_1980}
Hutson V and Pym J~S 1980 {\em Applications of functional analysis and operator
  theory\/} ({\em Mathematics in science and engineering\/} no no. 146) (London
  ; New York: Academic Press) ISBN 978-0-12-363260-9

\bibitem{gbur_2011}
Gbur G~J 2011 {\em Mathematical Methods for Optical Physics and Engineering\/}
  (Cambridge University Press)

\bibitem{schuster_hybrid_1985}
Schuster G~T 1985 {\em The Journal of the Acoustical Society of America\/} {\bf
  77} 865--879 ISSN 0001-4966
  \urlprefix\url{http://asa.scitation.org/doi/10.1121/1.392055}

\bibitem{martin_multiple_2006}
Martin P~A 2006 {\em Multiple scattering: interaction of time-harmonic waves
  with {N} obstacles\/} ({\em Encyclopedia of mathematics and its
  applications\/} no 107) (Cambridge ; New York: Cambridge University Press)
  ISBN 978-0-521-86554-8 oCLC: ocm70059806

\bibitem{bourlier_sub-domain_2015}
Bourlier C, Bellez S, Li H and Kubicke G 2015 {\em IEEE Trans. Antennas
  Propagat.\/} {\bf 63} 659--666 ISSN 0018-926X, 1558-2221
  \urlprefix\url{http://ieeexplore.ieee.org/document/6965626/}

\bibitem{mishchenko_independent_2018}
Mishchenko M~I 2018 {\em OSA Continuum\/} {\bf 1} 243 ISSN 2578-7519
  \urlprefix\url{https://www.osapublishing.org/abstract.cfm?URI=osac-1-1-243}

\bibitem{mishchenko_multiple_2008}
Mishchenko M~I 2008 {\em Rev. Geophys.\/} {\bf 46} RG2003 ISSN 8755-1209
  \urlprefix\url{http://doi.wiley.com/10.1029/2007RG000230}

\bibitem{heidinger_successive-order--interaction_2006}
Heidinger A~K, O’Dell C, Bennartz R and Greenwald T 2006 {\em J. Appl.
  Meteor. Climatol.\/} {\bf 45} 1388--1402 ISSN 1558-8424, 1558-8432
  \urlprefix\url{http://journals.ametsoc.org/doi/10.1175/JAM2387.1}

\bibitem{burkholder_forward-backward_2005}
Burkholder R and Lundin T 2005 {\em IEEE Trans. Antennas Propagat.\/} {\bf 53}
  793--799 ISSN 0018-926X, 1558-2221
  \urlprefix\url{http://ieeexplore.ieee.org/document/1391151/}

\bibitem{gershenzon_paper_2018}
{Gershenzon} I, {Brick} Y and {Boag} A 2018 {\em IEEE Transactions on Antennas
  and Propagation\/} {\bf 66} 871--883

\bibitem{obelleiro_1995}
{Obelleiro-Basteiro} F, {Luis Rodriguez} J and {Burkholder} R~J 1995 {\em IEEE
  Transactions on Antennas and Propagation\/} {\bf 43} 356--361

\bibitem{renardy_introduction_2004}
Renardy M and Rogers R~C 2004 {\em An introduction to partial differential
  equations\/} 2nd ed ({\em Texts in applied mathematics\/} no~13) (New York:
  Springer) ISBN 978-0-387-00444-0

\bibitem{menchon-born}
Lopez-Menchon H, Rius J~M, Heldring A and Ubeda E 2021 {\em IEEE Transactions
  on Antennas and Propagation\/}  1--1

\bibitem{sadiku_monte_nodate}
Sadiku M~N~O   240

\bibitem{noebauer_monte_2019}
Noebauer U~M and Sim S~A 2019 {\em Living Rev Comput Astrophys\/} {\bf 5} 1
  ISSN 2367-3621, 2365-0524
  \urlprefix\url{http://link.springer.com/10.1007/s41115-019-0004-9}

\bibitem{barker_monte_2003}
Barker H~W, Goldstein R~K and Stevens D~E 2003 {\em JOURNAL OF THE ATMOSPHERIC
  SCIENCES\/} {\bf 60} 14

\bibitem{deutschmann_monte_2011}
Deutschmann T, Beirle S, Frieß U, Grzegorski M, Kern C, Kritten L, Platt U,
  Prados-Román C, Pukite J, Wagner T, Werner B and Pfeilsticker K 2011 {\em
  Journal of Quantitative Spectroscopy and Radiative Transfer\/} {\bf 112}
  1119--1137 ISSN 00224073
  \urlprefix\url{https://linkinghub.elsevier.com/retrieve/pii/S0022407310004668}

\bibitem{ji_convergence_2013}
Ji H, Mascagni M and Li Y 2013 {\em SIAM J. Numer. Anal.\/} {\bf 51} 2107--2122
  ISSN 0036-1429, 1095-7170
  \urlprefix\url{http://epubs.siam.org/doi/10.1137/130904867}

\bibitem{stein_complex_2003}
Stein E~M and Shakarchi R 2003 {\em Complex analysis\/} ({\em Princeton
  lectures in analysis\/} no~2) (Princeton, N.J: Princeton University Press)
  ISBN 978-0-691-11385-2 oCLC: ocm51738532

\bibitem{lang_complex_1999}
Lang S 1999 {\em Complex {Analysis}\/} ({\em Graduate {Texts} in
  {Mathematics}\/} vol 103) (New York, NY: Springer New York) ISBN
  978-1-4419-3135-1 978-1-4757-3083-8
  \urlprefix\url{http://link.springer.com/10.1007/978-1-4757-3083-8}

\bibitem{kythe_handbook_2019}
Kythe P~K 2019 {\em Handbook of conformal mappings and applications\/} (Boca
  Raton: CRC Press, Taylor \& Francis Group) ISBN 978-1-315-18023-6
  978-1-351-71872-1

\bibitem{asmar_complex_2018}
Asmar N~H and Grafakos L 2018 {\em Complex {Analysis} with {Applications}\/}
  Undergraduate {Texts} in {Mathematics} (Cham: Springer International
  Publishing) ISBN 978-3-319-94062-5 978-3-319-94063-2
  \urlprefix\url{http://link.springer.com/10.1007/978-3-319-94063-2}

\bibitem{starke_hybrid_1993}
Starke G and Varga R~S 1993 {\em Numer. Math.\/} {\bf 64} 213--240 ISSN
  0029-599X, 0945-3245
  \urlprefix\url{http://link.springer.com/10.1007/BF01388688}

\bibitem{driscoll_schwarz-christoffel_2002}
Driscoll T~A and Trefethen L~N 2002 {\em Schwarz-{Christoffel} mapping\/} ({\em
  Cambridge monographs on applied and computational mathematics\/} no v. 8)
  (Cambridge ; New York: Cambridge University Press) ISBN 978-0-521-80726-5

\bibitem{laura-conformal-mapping}
Schinzinger R 2003 Conformal mapping : methods and applications

\bibitem{hanson-yakovlev}
Hanson G~W 2002 {\em Operator theory for electromagnetics : an introduction\/}
  (New York: Springer) ISBN 9781441929341

\bibitem{saad_iterative_2003}
Saad Y 2003 {\em Iterative Methods for Sparse Linear Systems\/} 2nd ed (USA:
  Society for Industrial and Applied Mathematics) ISBN 0898715342

\bibitem{thehee_generalization_2002}
Thaheem A~B and Laradji A 2003 {\em International Journal of Mathematical
  Education in Science and Technology\/} {\bf 34} 905--907 (\textit{Preprint}
  \eprint{https://doi.org/10.1080/00207390310001595410})
  \urlprefix\url{https://doi.org/10.1080/00207390310001595410}

\bibitem{hackbusch_iterative_2016}
Hackbusch W 2016 {\em Iterative {Solution} of {Large} {Sparse} {Systems} of
  {Equations}\/} ({\em Applied {Mathematical} {Sciences}\/} vol~95) (Cham:
  Springer International Publishing) ISBN 978-3-319-28481-1 978-3-319-28483-5
  \urlprefix\url{http://link.springer.com/10.1007/978-3-319-28483-5}

\bibitem{vanbladel2007electromagnetic}
Van~Bladel J 2007 {\em Electromagnetic Fields\/} IEEE Press Series on
  Electromagnetic Wave Theory (Wiley) ISBN 9780470124574
  \urlprefix\url{https://books.google.am/books?id=bupYviuRMLgC}

\bibitem{harrington}
Harrington R~F 1993 {\em Field Computation by Moment Methods\/} (Wiley-IEEE
  Press) ISBN 0780310144

\bibitem{rahola_eigenvalues_2000}
Rahola J 2000 {\em SIAM J. Sci. Comput.\/} {\bf 21} 1740--1754 ISSN 1064-8275,
  1095-7197 \urlprefix\url{http://epubs.siam.org/doi/10.1137/S1064827598338962}

\bibitem{chew_fft}
Gan H and Chew W 1995 {\em Journal of Electromagnetic Waves and Applications\/}
  {\bf 9} 1339--1357 (\textit{Preprint}
  \eprint{https://www.tandfonline.com/doi/pdf/10.1163/156939395X00082})
  \urlprefix\url{https://www.tandfonline.com/doi/abs/10.1163/156939395X00082}

\bibitem{evans_MC}
Evans M and Swartz T 2000 {\em Approximating Integrals via Monte Carlo and
  Deterministic Methods\/}

\bibitem{leobacher_introduction_2014}
Leobacher G and Pillichshammer F 2014 {\em Introduction to {Quasi}-{Monte}
  {Carlo} {Integration} and {Applications}\/} Compact {Textbooks} in
  {Mathematics} (Cham: Springer International Publishing) ISBN
  978-3-319-03424-9 978-3-319-03425-6
  \urlprefix\url{http://link.springer.com/10.1007/978-3-319-03425-6}

\bibitem{balanis2012advanced}
Balanis C 2012 {\em Advanced Engineering Electromagnetics, 2nd Edition\/} (New
  York: Wiley)

\bibitem{dimov_book}
Dimov I~T and McKee S 2004 {\em Monte Carlo Methods for Applied Scientists\/}
  (World Scientific Press) ISBN 9810223293

\bibitem{sertel_integral_2012}
{K Sertel} and {J Volakis} 2012 {\em Integral {Equation} {Methods} for
  {Electromagnetics}\/} (Institution of Engineering and Technology) ISBN
  978-1-891121-93-7 978-1-61353-112-9
  \urlprefix\url{https://digital-library.theiet.org/content/books/ew/sbew045e}

\bibitem{zwamborn}
Zwamborn P and van~den Berg P 1992 {\em IEEE Transactions on Microwave Theory
  and Techniques\/} {\bf 40} 1757--1766

\bibitem{Czechowski}
Czechowski K, Battaglino C, McClanahan C, Iyer K, Yeung P~K and Vuduc R~W 2012
  On the communication complexity of 3d ffts and its implications for exascale.
  {\em ICS\/} ed Banerjee U, Gallivan K~A, Bilardi G and Katevenis M (ACM) pp
  205--214 ISBN 978-1-4503-1316-2
  \urlprefix\url{http://dblp.uni-trier.de/db/conf/ics/ics2012.html#CzechowskiBMIYV12}

\bibitem{Solihin}
Solihin Y 2015 {\em Fundamentals of Parallel Multicore Architecture\/} 1st ed
  (Chapman and Hall/CRC) ISBN 1482211181

\bibitem{Grama}
Kumar V, Grama A, Gupta A and Karypis G 1994 {\em Introduction to Parallel
  Computing: Design and Analysis of Algorithms\/} (USA: Benjamin-Cummings
  Publishing Co., Inc.) ISBN 0805331700

\bibitem{Yavits}
Yavits L, Morad A and Ginosar R 2014 {\em Parallel Computing\/} {\bf 40} 1--16
  ISSN 0167-8191
  \urlprefix\url{https://www.sciencedirect.com/science/article/pii/S0167819113001324}

\bibitem{ayala}
Ayala A, Tomov S, Luo X, Shaeik H, Haidar A, Bosilca G and Dongarra J 2019
  Impacts of multi-gpu mpi collective communications on large fft computation
  {\em 2019 IEEE/ACM Workshop on Exascale MPI (ExaMPI)\/} pp 12--18

\end{thebibliography}

\end{document}